\documentclass[aps,pra,twocolumn,floatfix]{revtex4-2}
\usepackage{blindtext}
\usepackage{bm}
\usepackage{braket}
\usepackage{hyperref}
\usepackage{amssymb}
\usepackage{amsmath}
\usepackage{graphicx}
\usepackage{float}
\usepackage{bbold}
\usepackage{color}
\usepackage{mathtools}
\usepackage{xr}

\begin{document}
\title{Super-resolution microscopy via fluctuation-enhanced spatial mode demultiplexing}
\author{Stanis{\l}aw Kurdzia{\l}ek}
\affiliation{Faculty of Physics, University of Warsaw, Pasteura 5, 02-093 Warszawa, Poland}
\affiliation{Department of Physics, University of Oxford, Parks Road, Oxford OX1 3PU, UK}
\begin{abstract}
  We introduce a superresolution technique that combines spatial mode demultiplexing (SPADE) with emitter blinking. We show that temporal fluctuations not only enhance the precision of SPADE imaging, but also drastically simplify the measurement required to recover full object information---in the presence of fluctuations, SPADE can be replaced by the much simpler image inversion interferometry. Both gains are enabled by exploiting temporal cumulants of the detected signal. We further show that the proposed fluctuation-enhanced techniques are significantly more robust to dark-count noise than conventional SPADE. 
\end{abstract}
\maketitle
\twocolumngrid
\section{Introduction}
Light diffraction limits the resolution of conventional light microscopy to $\sim 200~\textrm{nm}$ (Rayleigh criterion \cite{Born2013}). Recently developed super-resolution methods \cite{Betzig1995, Betzig2006, Rust2006, Hell2007, Dertinger2009, Bonnie2011, Monticone2014, Moerner2015,  Tsang_2016, Paur2016, Yang2016, Tham2017, Parniak2018, Erkmen2008, Boto2000, Taylor2014, Rozema2014, Dowling2015, Genovese2016, Schnell2019, Gustafsson2000, Muller2010, Schwarz2012, schwartz2013superresolution,Fujita2007, Yamanaka2008} overcome this barrier and, under certain conditions, reveal details at nanometers scale\cite{Balzarotti2017, Moosmayer2024, Wang2024}.

Traditional imaging assumes (i) uniform and classical illumination, (ii) a linear, time-invariant and classical response of the imaged object, and (iii) measurement of output field with a camera placed in an image plane. Super-resolution techniques break at least one of assumptions (i)–(iii), yielding three broad classes.

Category (i) covers structured illumination microscopy (SIM) \cite{Gustafsson2000} or image scanning microscopy (ISM) \cite{Muller2010}, as well as approaches with nonclassical light sources \cite{Erkmen2008, Taylor2014, Rozema2014}. 

Category (ii) exploits emitter dynamics \cite{Rust2006, Dertinger2009, Betzig2006}, non-Poissonian photon statistics \cite{Schwarz2012,schwartz2013superresolution,Picariello2025}, or nonlinearity \cite{Fujita2007, Yamanaka2008}. Stochastic optical reconstruction microscopy (STORM)\cite{Rust2006} and photoactivated localization microscopy (PALM) \cite{Betzig2006}, both awarded Nobel Prize in Chemistry in 2014, exploit \textit{blinking}: under uniform excitation, fluorophores randomly switch on and off, leaving only a few active at a time, which enables precise localization. Stochastic optical fluctuation imaging (SOFI) \cite{Dertinger2009, Dertinger2010} also uses blinking but does not require isolating single emitters; instead, it analyzes temporal cumulants of image sequences. Photon antibunching, which yields sub-Poissonian statistics, can likewise enable super-resolution imaging \cite{Schwarz2012, schwartz2013superresolution}.

%Notably, STORM and PALM received the 2014 Nobel Prize in Chemistry. 
%Stimulated emission depletion (STED)---also Nobel-prize winning---fits both (i) and (ii): two beams are used, one to excite fluorescence and another to deplete it, which allows to make an active area far below diffraction limit.

Category (iii) is the youngest family of techniques, inspired by Tsang et al.~\cite{Tsang_2016}, who showed that resolution can be enhanced by replacing intensity measurements with more informative detection of the optical field. In particular, measuring in the Hermite–Gaussian basis is optimal for resolving two closely spaced point sources. This idea is realized by spatial mode demultiplexing (SPADE) measurement, which we use to refer to Tsang’s imaging scheme. A closely related, less powerful but often more accessible variant is image inversion interferometry (III) \cite{Smith2005,Sandeau2006, Nair2016}, which sorts the field into even and odd spatial modes [this technique is also sometimes called Super Localization by Image inVERsion (SLIVER)].  Superresolution can also be achieved via homodyne or heterodyne detection of the image-plane field \cite{Yang2016, Yang2017, Datta2020} or by using Hong-Ou-Mandel effect \cite{Parniak2018}.

Several techniques combine categories (i) and (ii), for example the Nobel Prize-winning stimulated emission depletion (STED) \cite{Hell2007} microscopy or stochastic optical fluctuation image scanning microscopy (SOFISM) \cite{Classen2017,Sroda2020}, which merges SOFI and ISM. Yet no method has so far combined the resolution gains of non-Poissonian sources [category (ii)] with measurement strategies of category (iii) such as SPADE.

In this work we introduce stochastic optical fluctuation SPADE (SOFSPADE), which for the first time combines emitter fluctuations with SPADE detection to realize joint resolution gain. Equally important, temporal fluctuations allow us to significantly simplify the measurement: instead of full SPADE, one can perform the much simpler III measurement while still recovering information about complex object structure. We will call this technique stochastic optical fluctuation image inversion interferometry (SOFIII). 

We derive a general estimator of object's spatial moments applicable to both techniques (SOFSPADE and SOFIII), and prove its near-optimality by comparing its precision with the fundamental Cram\'er-Rao (C-R) bound. Numerically, we find that the advantage from fluctuations grows with the moment order: for example, in the absence of technical noise, SOFSPADE reaches the same precision for the 8th spatial moment as conventional SPADE using nearly 1000 times fewer frames. This advantage grows dramatically, to $10^8$-fold, once dark-count noise is present.

The paper is organized as follows. In Sec.~\ref{sec:opt_imaging_models} we introduce general formalism describing linear optical imaging, and discuss direct imaging (DI) and known superresolution techniques such as SPADE, III, interferometric SPADE (iSPADE) and SOFI using this formalism. Section~\ref{sec:SOFSPADE} contains the main result of this work---the working principle of novel SOFSPADE and SOFIII superresolution techniques, with practical recipe for object's spatial moments estimator. Then, in Sec.~\ref{sec:simulation}, we show numerical results demonstrating the practical advantage from fluctuations and the optimality of proposed estimator, also in the presence of dark-count noise (Sec.~\ref{sec:dark_counts}).
%Additional practical and theoretical aspects of proposed techniques are discussed in Sec.~\ref{sec:practical} and Sec.~\ref{sec:theoretical} respectively, and Sec.~\ref{sec:discussion} contains the discussion of the results.
Additional practical aspects of proposed techniques are discussed in Sec.~\ref{sec:practical}, Sec.~\ref{sec:discussion} contains the summary and the discussion of the results. 

\section{Optical imaging models}
%\subsection{Direct imaging (DI)}
\label{sec:opt_imaging_models}
Any linear imaging system is characterized by a transfer function $T(j|\boldsymbol{r})$, which is a probability of detecting a photon emitted from object's position $\boldsymbol{r}$ at detector labeled by $j$. Consider an object consisting of multiple independent, incoherent, weak ($\ll 1$ photons per temporal mode) sources with positions $\boldsymbol{r}_i$ and constant brightnesses $q_i$. Each source emits $n_i \sim \textrm{Poiss}(q_i)$ photons per unit time, where $\textrm{Poiss}(\mu)$ denotes Poissonian random variable with mean $\mu$. The detectors' photon counts $n_{j}$ are then independent and
\begin{equation}
\label{eq:poiss_gen}
    n_{j} \sim \textrm{Poiss}(I_{j}), \quad I_{j} = \sum_i T(j|\boldsymbol{r}_i) q_i,
\end{equation}
where $I_{j}$ are detector intensities. Note that we cannot measure $I_{j}$ exactly, we only have access to shot-noise-affected photon counts $n_{j}$.
In direct imaging (DI), detectors correspond to camera pixels at positions $\boldsymbol{r}_j$. The resulting transfer function is \begin{equation}
T^\textrm{DI}(\boldsymbol{r}_j|\boldsymbol{r}) = \eta U(\boldsymbol{r}_j - \boldsymbol{r}),
\end{equation}
where $U$ is the point spread function (PSF). Coordinates were rescaled such that system magnification is unity, and $\eta$ denotes signal attenuation factor. From now on we set $\eta=1$, so that $q_i$ represent total brightnesses observed in the image plane rather than at the object.

 We also assume a 1D object, and write positions as scalars $x$ instead of vectors $\boldsymbol{r}$.  This simplification highlights the core ideas of introduced imaging techniques, all the results extend to 2D, as we will show in Sec.~\ref{sec:2D}. Furthermore, we set a Gaussian PSF
 $
    U(x) = \frac{1}{\sqrt{2 \pi \sigma^2}}e^{-\frac{x^2}{2 \sigma^2}},
$
extensions to more realistic PSFs are also discussed in Sec.~\ref{sec:2D}.

\subsection{Spatial mode demultiplexing imaging (SPADE)}

\begin{figure}
    \centering
    \includegraphics[width = 0.75\columnwidth]{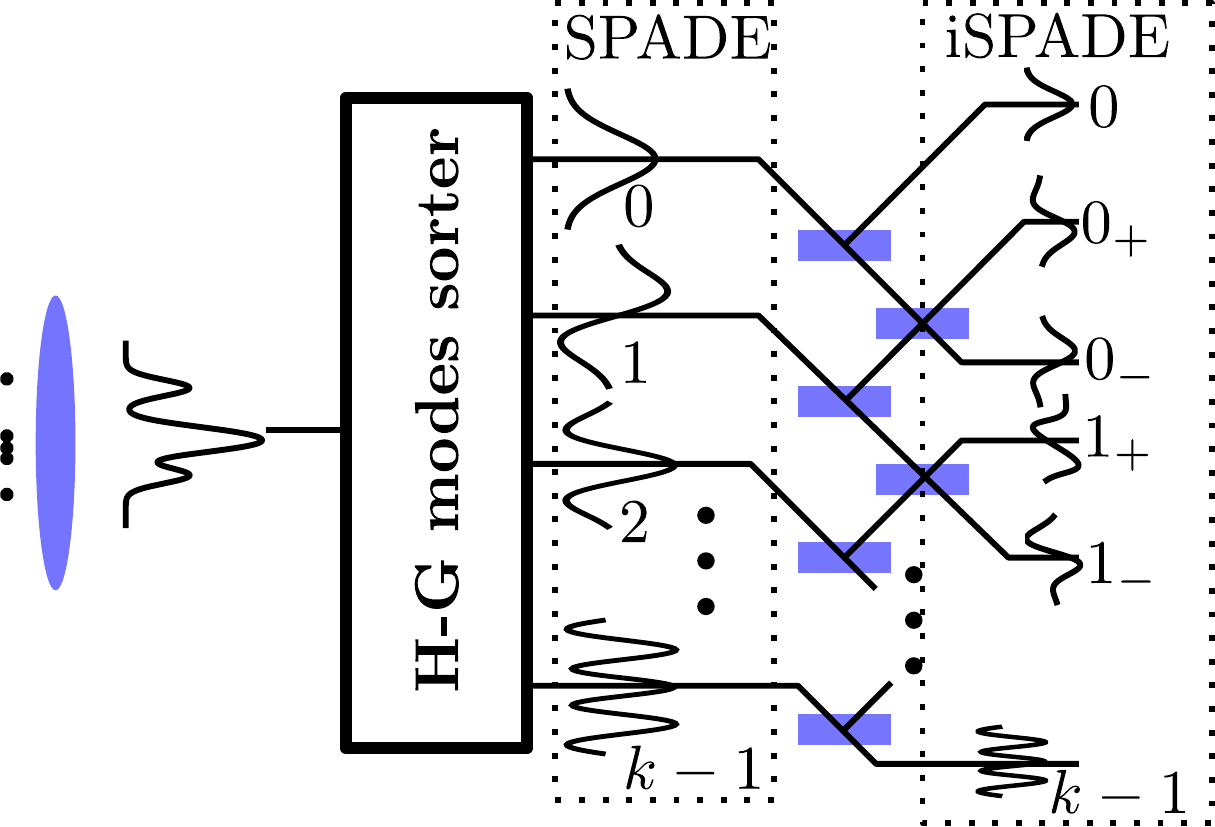}
    \caption{In SPADE technique, image plane field is decomposed into $k$ Hermite-Gaussian (H-G) modes, which allows us to estimate lowest $k$ object's even spatial moments. To estimate odd spatial moments, one needs to interfere neighbouring H-G modes, which leads to iSPADE technique. }
    \label{fig:iSPADE}
\end{figure}
In SPADE imaging technique, we replace the position measurement with the measurement in Hermite-Gaussian (H-G) modes basis \cite{Tsang_2016} 
%$\phi_j(x)$, 
\begin{equation}
\label{eq:HG_modes}
\phi_j(x') = \frac{1}{\sqrt[4]{2 \pi \sigma^2}} \frac{1}{\sqrt{2^j j!}} H_j \left(\frac{x'}{\sigma \sqrt{2}}\right) \exp \left( - \frac{x'^2}{4 \sigma^2} \right),
\end{equation}
$j \in \{0,1,2,...\}$, $H_j$ are Hermite polynomials, $x'$ is a position in the image plane. This can be implemented by commercially available multi-plane light converters \cite{Tan2023Optica,Rouviere2024, Santamaria2024} or by heterodyning image field with local oscillators with varying space profiles \cite{Yang2016,Pushkina2021, Frank2023Optica, Duplinskiy2025}---this latter method, however, causes additional signal losses.
%, $H_j$ are Hermite polynomials.
%defined by generating function
%\begin{equation}
%    e^{2xt - t^2} = \sum_{j=0}^\infty H_j(x) \frac{t^j}{j!}.
%\end{equation}
The resulting transfer function is
\begin{equation}
\label{eq:trasfer_SPADE}
    T^\textrm{S}(j|x) = \frac{1}{4^j j !} e^{-\frac{x^2}{4 \sigma^2}} \left( \frac{x}{\sigma} \right)^{2j},
\end{equation}
where $x$ is measured respectively to the center of a mode sorting device. Notably, $T(0|0) = 1$, $T(j|0) = 0$ for $j>0$, so all H-G modes except $\phi_0$ are dark for a point source at $x=0$. Small changes of position and shape then produces signal in these initially dark modes. This is the reason for resolution enhancement achieved by SPADE---it is easier to distinguish zero from nonzero than to resolve small changes atop a large background. This mechanism can be used for exoplanet detection\cite{Huang2021}---higher H-G modes are dark for a lone star but become excited for a star with an orbiting planet. 

The superresolving power of SPADE can be demonstrated using a minimal example of two identical sources 
%with $x_1 = \frac{s}{2}$, $x_2 = -\frac{s}{2}$ 
of unknown separation $s$ and a known centroid. For DI, the mean squared error (MSE) of any unbiased separation estimator $\hat s$,  satisfies, for a fixed photon budget, $\delta^2 \hat s  = \langle (s-\hat s)^2\rangle \overset{s \rightarrow 0}{\longrightarrow} \infty$---this is called \textit{Rayleigh's curse} \cite{Tsang_2016}.
%this is called a Rayleigh's curse, since  $\Delta^2 \hat s$ starts to grow rapidly for $s$ below Rayleigh's resolution limit.
Surprisingly, with SPADE one can construct an estimator for which $\delta^2 \hat s$ remains finite and constant even for $s \rightarrow0$, enabling precise separation estimation even far below Rayleigh's limit. While noise and mode sorter misalignment reduce SPADE precision \cite{zhou2019, Oh2021}, it still outperforms DI in sub-Rayleigh regime, which was demonstrated both theoretically \cite{Len2020,Lupo2020,Gessner2020, Almeida2021} and experimentally \cite{Tan2023Optica, Rouviere2024, Santamaria2024}.

After Taylor expanding \eqref{eq:trasfer_SPADE} and inserting it into \eqref{eq:poiss_gen}, we get light intensities on SPADE detectors for a general multi-emitter object,
\begin{equation}
\label{eq:SPADE_Ij}
    I_j^{\textrm{S}} = \sum_{\mu = j}^\infty A_{\mu,j}^{\textrm{S}} \theta_{2 \mu},\quad A_{\mu,j}^\textrm{S} = \frac{(-1)^{\mu-j}}{4^\mu j! (\mu-j)!},
%= \frac{1}{4^j j!} \theta_{2j} + \mathcal{O}(\Delta^{2j+2})
\end{equation}
%\begin{equation}
%\label{eq:SPADE_Ij}
%    I_j^{(\textrm{S})} = \sum_{\mu = j}^\infty A^\mu_j \theta_{2 \mu},\quad A^\mu_j = \frac{(-1)^{\mu-j}}{4^\mu j! (\mu-j)!},
%= \frac{1}{4^j j!} \theta_{2j} + \mathcal{O}(\Delta^{2j+2})
%\end{equation}
where $\theta_\mu = \sum_i q_i (x_i/\sigma)^\mu$ are object's dimensionless spatial moments.
%Let us now consider SPADE imaging of a more complex object consisting of multiple Poissonian sources with positions $x_i$ and powers $q_i$; then, the intensity of light measured in $\phi_j$ is a linear combination of 
%\begin{equation}
 %   I_j = \sum_i q_i T(j|x_i),
%\end{equation}
%the corresponding photon numbers are Poissonian random variables $n_j \sim \textrm{Poiss}(I_j)$. 

%After Taylor expanding \eqref{} and inserting it into \eqref{}, we express $I_j$ as a linear combination of an object's dimensionless spatial moments $\theta_\mu = \sum_i q_i (x_i/\sigma)^i$;

Consequently, object's even spatial moments can be reconstructed  by performing SPADE measurement and inverting linear relation \eqref{eq:SPADE_Ij}. In a subdiffraction limit $\Delta = \max_i |x_i/\sigma| \ll 1$,  \eqref{eq:SPADE_Ij} reduces to \cite{Tsang_2017}
\begin{equation}
   I_j^\textrm{S} = \frac{1}{4^j j!} \theta_{2j} + \mathcal{O}(\Delta^{2j+2}),
\end{equation}
so $2j$th moment estimator is simply $\hat \theta_{2j} = 4^j j! n^{\textrm{S}}_{j}$.
%Therefore, for very small objects, $I_j$ is proportional to $2j$-th spatial moment of an object, which makes moments estimation straightforward. 

Unfortunately, SPADE only makes it possible to estimate even moments because $I_j^{\textrm{S}}$ do not depend on odd moments. To remedy this, interferometric SPADE (iSPADE) technique was introduced \cite{Tsang_2017}, in which a mode sorting device is followed by beam splitters mixing neighbouring modes, as shown in Fig.~\ref{fig:iSPADE}.
When the SPADE sorter distinguishes $k$ modes $\phi_0, \phi_1,...,\phi_{k-1
}$, then there are $2k$ iSPADE outputs labeled by $0, j _\pm$ for $j \in \{0,1,...,k-2\}$, $k-1$; outputs $j_\pm$ contain interfered modes $j$ and $j+1$. The corresponding transfer functions are 
\begin{align}
    T^{\textrm{iS}}(j|x) &= \frac{1}{2} T^{\textrm{S}}(j|x)~~ \textrm{for}~~ j \in \{0,k\} \\
    T^{\textrm{iS}}(j_\pm|x) &= \frac{1}{4^{j+1} j!} e^{-\frac{x^2}{4 \sigma^2}} \left( \frac{x}{\sigma} \right)^{2j} \left( 1 \pm \frac{1}{2\sqrt{j+1}} \frac{x}{\sigma}  \right)^2.
\end{align}

These transfer functions depend on odd powers of $x$,  which allows us to estimate odd moments---in particular, in the subdiffraction limit \cite{Tsang_2017}
\begin{align}
    I_{j_+}^\textrm{iS} + I_{j_-}^\textrm{iS} &= \frac{1}{2} \frac{1}{4^j j!} \theta_{2j} + \mathcal{O}(\Delta^{2j+2}),\\
    I_{j_+}^\textrm{iS} - I_{j_-}^\textrm{iS} &= \frac{1}{2} \frac{1}{4^j j! \sqrt{j+1}} \theta_{2j+1} + \mathcal{O}(\Delta^{2j+3}),
\end{align}
so $\hat \theta_{2j} = 2\cdot 4^j j! (n_{j_+}^\textrm{iS} + n_{j_-}^\textrm{iS})$, $\hat \theta_{2j+1} = 2\cdot 4^j j! \sqrt{j+1} (n_{j_+}^\textrm{iS} - n_{j_-}^\textrm{iS})$.

Relative errors of  moments estimation for SPADE and iSPADE techniques satisfy \cite{Tsang_2017, Tsang2018} $\delta \hat \theta_\mu/\theta_\mu = \mathcal{O}(\Delta^{-\lceil \frac{\mu}{2}\rceil})$ ---this is the optimal scaling with $\Delta$ over all measurements allowed by quantum mechanics\cite{Tsang_2019, Tsang_2019_Semiparametric, Tsang2021}. For DI, $\delta \hat \theta_\mu/\theta_\mu = \mathcal{O}(\Delta^{-\mu})$ \cite{Tsang_2017}, so (i)SPADE gives a better scaling for $\mu \ge 2$. While these results are derived for $\Delta \rightarrow 0$, numerical studies confirm (i)SPADE's optimality in a more realistic $\Delta \approx 1$ regime \cite{tan2023quantum}.

In principle, iSPADE enables any object reconstruction by estimating spatial moments and expressing intensity distribution in terms of them. This was demonstrated in a proof-of-principle experiment, with neural networks assisting in processing noisy data \cite{Pushkina2021,Frank2023Optica} . 

However, this approach is impractical for objects much larger than the PSF ($\Delta \gg 1$) since large number of moments is needed, and the advantage of (i)SPADE over DI is not clear in this regime~\cite{Wang2023}.  It is therefore much more promising to use (i)SPADE in scanning, confocal microscopy, where the illuminated part of an object is of the size of diffraction spot, so $\Delta \approx 1$.
The object can then be scanned, (i)SPADE measurements taken locally, and the full image reconstructed from multiple spatial moment maps using generalized Richardson-Lucy deconvolution algorithm \cite{Bearne2021} or neural networks \cite{Zhang2025}. 

Applying (i)SPADE in practical microscopy with fluorescent emitters and high-numerical-aperture (NA) objectives  is technically and conceptually demanding \cite{Greenwood2023}. A simpler and much easier to implement \cite{Larson2021, Aiello2025} alternative is image inversion interferometry (III), which has already demonstrated superresolution of two fluorescent emitters in a high-NA setup \cite{Mitchell2024}. III sorts the optical field into even and odd modes by interfering the image with its spatial inversion. The corresponding transfer function is 
\begin{equation}
    T^{\textrm{III}}(\pm|x) = \frac{1}{2} \left(1 \pm e^{-\frac{x^2}{2 \sigma ^2}} \right),
\end{equation}
$+$/$-$ correspond to even/odd modes.
For two point sources, III and SPADE achieve the same precision for $s \rightarrow 0$ \cite{Nair2016}; however, III becomes worse for larger separations. More importantly, III provides limited information about complex objects; with only two outcomes, only two independent object properties can be inferred. In the subdiffraction regime, intensities of even and odd modes are
\begin{equation}
    I^{\textrm{III}}_+ = \theta_0 + \mathcal{O}(\Delta^2), \quad I^{\textrm{III}}_- = \frac{\theta_2}{4} + \mathcal{O}(\Delta^4),
\end{equation}
so only $0$th and 2nd spatial moments can be estimated.

An important result of this work, demonstrated in Sec.~\ref{sec:gen_princ}, is that temporal fluctuations enable extraction of all even moments from III, providing the full information of SPADE with a much simpler measurement.

\subsection{Stochastic optical fluctuation imaging (SOFI)}
\label{sec:SOFI}
Let us consider super-Poissonian sources with time-fluctuating brightnesses, such as quantum dots \cite{efros2016origin} or dyes \cite{dickson1997off} used in fluorescence microscopy. 
 We divide observation time into \textit{frames}. In each frame, the photon number from $i$th source follows $n_i \sim \textrm{Poiss}(q_i)$, where  $q_i$ is a random variable representing the frame-integrated source brightness. Thus, fluctuations of $n_i$ arise both from variability of $q_i$ and and from shot noise. For multiple sources at positions $x_i$, the pixel intensity for DI is $I(x_j) = \sum_i q_i U(x_j-x_i)$, which also fluctuates in time. Recording multiple frames yields different samples of a random variable $I(x_j)$.

SOFI \cite{Dertinger2009, Dertinger2010} exploits these fluctuations to enhance resolution. 
The $r$th cumulant of a random variable $X$ is defined as 
\begin{equation}
\label{eq:cum_single}
    \kappa^{(r)}(X) =\left. \frac{\textrm{d}}{\textrm{d}t^r} \right|_{t=0} \log \langle e^{tX} \rangle,
\end{equation}
for example, $\kappa^{(1)}(X) = \langle X \rangle$,  $\kappa^{(2)}(X) = \langle X^2 \rangle - \langle X \rangle^2$, $\langle \bullet \rangle$ denotes expected value.
For independent $q_j$s
\begin{equation}
\label{eq:SOFI}
    \kappa^{(r)} (I(x_j)) = \sum_i \kappa^{(r)}(q_i) U^r(x_j - x_i).
\end{equation}
This is the core equation of SOFI---in the $r$th cumulant, the PSF is effectively replaced with its $r$th power, and consequently narrowed by a factor $\sqrt r$. In principle, one can arbitrarily narrow the PSF using cumulants of large orders; however, in practice higher cumulants become very noisy, which limits the resolution gain \cite{kurdzialek2021super}. 

Importantly, in practice one measures shot noise-affected photon counts $n(x_j)$, not the noiseless intensities $I(x_j)$, and $\kappa^{(r)}(I(x_j)) \ne \kappa^{(r)}(n(x_j))$ for $r \ge 2$. Therefore, \eqref{eq:SOFI} does not hold for experimentally measured cumulants---the discrepancy is especially significant at low photon numbers. This issue was recently addressed in Ref.~\cite{Picariello2025}, which showed that resolution enhancement by $\sqrt{r}$  requires using  an appropriate linear combination of measured cumulants $\kappa^{(1)}$, $\kappa^{(2)},..., \kappa^{(r)}$ rather than $\kappa^{(r)}$ alone. We will return to this point in Sec.~\ref{sec:est_construction} when discussing the techniques introduced in the present work.

\section{Connecting SPADE and SOFI}
\label{sec:SOFSPADE}
\begin{figure*}[t]
\includegraphics[width = 0.85 \linewidth]{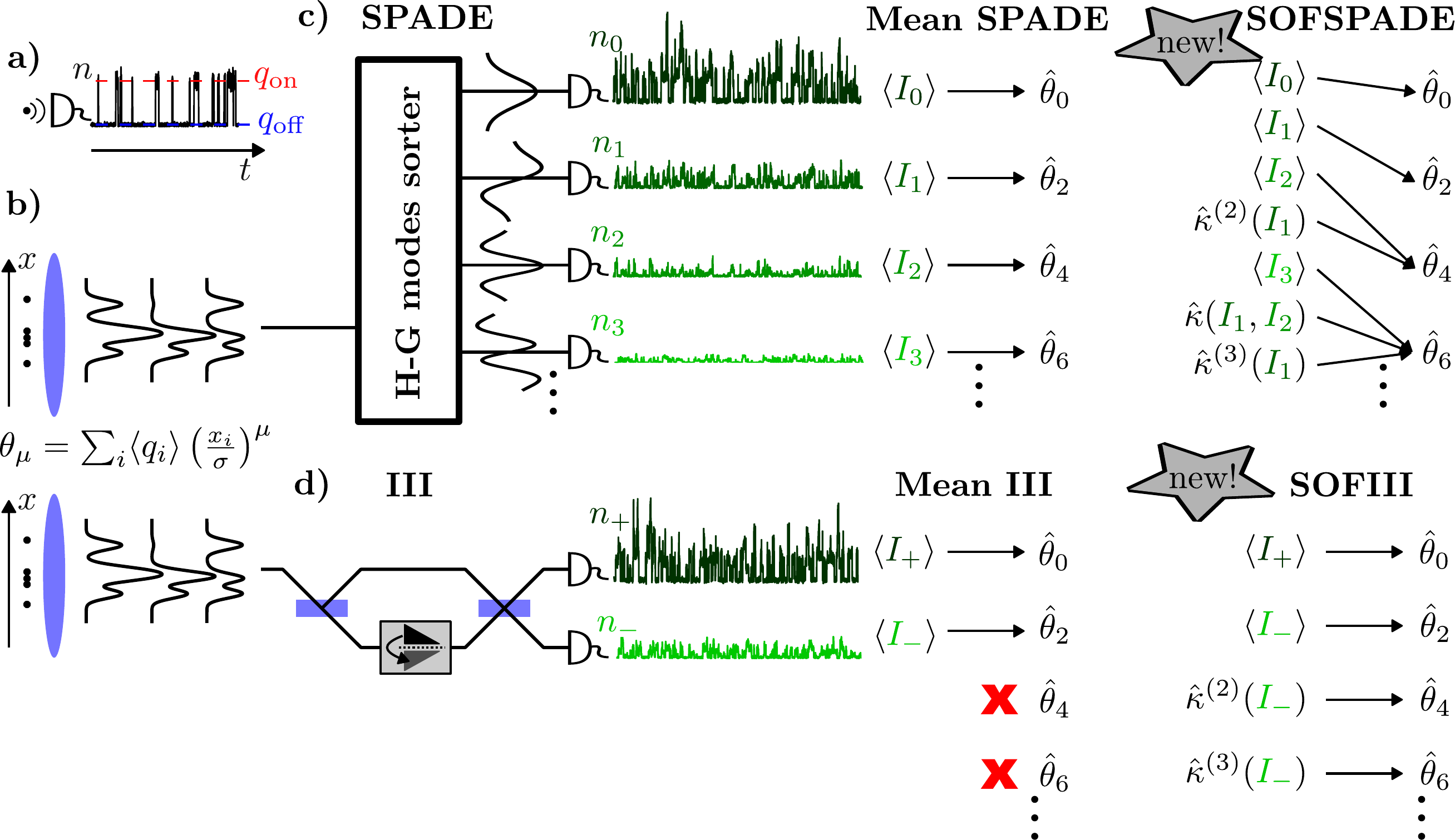}
\caption{Photon emission from quantum dots or dyes used to label an object fluctuates in time---typically, an emitter switches between two states characterized by brightnesses $q_\textrm{off}$ and $q_\textrm{on}$ (a). The resulting image also varies over time (b). In SPADE, the image plane field is sorted into H-G modes (c). 
Fluctuations of photon counts $n_j$ in different modes provide additional information about  object's higher ($\mu \ge 4$) spatial moments through suitable intensity cumulants---this leads to a proposed SOFSPADE technique. 
In III, the image-plane field is interfered with its spatial inversion (d), yielding a much simpler setup than SPADE, but providing access only to two outcomes corresponding to $0$th and $2$nd spatial moments.  
However, temporal fluctuations allow all even moments to be recovered 
from temporal cumulants of the odd-mode intensity $I_-$, 
which forms the basis of the proposed SOFIII technique.}
\label{fig:FESPADE_principle}
\end{figure*}
\subsection{General principle}
\label{sec:gen_princ}
It is natural to ask whether intensity fluctuations help to enhance resolution beyond DI measurement. We answer this question positively and introduce SOFSPADE technique which combines resolution gains from SOFI and SPADE.

Let us consider a 1D object consisting of independent fluctuating emitters imaged by a general linear system with a transfer function $T(j|x)$---\eqref{eq:poiss_gen} then applies with $q_i$ and $I_j$ interpreted as random variables.

Emitter fluctuations introduce correlations between detectors, which can be characterized by joint cumulants, defined as
\begin{equation}
\label{eq:cum_multi}
    \kappa(X_1, ...,X_n) = \left. \frac{\textrm{d}^r}{\textrm{d}t_1 \textrm{d}t_2...\textrm{d}t_r} \right|_{t_1,...,t_r = 0} \log \langle e^{\sum_i t_i X_i} \rangle,
\end{equation}
for example $\kappa(X_1, X_2) = \langle X_1 X_2\rangle - \langle X_1 \rangle \langle X_2 \rangle$ is a covariance. The general formula for multivariate cumulants is
\begin{equation}
\label{eq:mom_to_cum}
   \kappa(X_1,...,X_r) = \sum_\pi \Omega_\pi \prod_{B \in \pi} \left\langle \prod_{i \in B} X_i \right\rangle,
\end{equation}
%\begin{equation}
%    \kappa(X_1,...,X_r) = \sum_\pi (|\pi|-1)! (-1)^{|\pi|-1} \prod_{B \in \pi} \mu(X_i: i \in B),
%\end{equation}
where the sum runs over all partitions $\pi$ of $\{1,...,r\}$ into non-empty blocks $B$, $\Omega_\pi = (|\pi|-1)! (-1)^{|\pi|-1}$, where $|\pi|$ denotes the number of blocks in a given partition; see Appendix~\ref{app:cumulants} for explicit formulas for low-order cases and further cumulants properties.
%for example, for $r=4$ there are $15$ different partitions $\pi$, one of them is $\pi = \{\{1,4\}, \{2\}, \{3\}\}$, for which $|\pi| = 3$ since $\pi$ has three elements $B = \{1,4\}$, $B=\{2\}$, $B = \{3\}$. 
We denote by $\kappa^{(r_1,...,r_l)}(X_1,...,X_l)$ the joint cumulant in which $X_i$ appears $r_i$ times.

%Using statistical independence of emitters and joint cumulants properties, we show that (see ... for a proof)
The joint cumulant of detector intensities is
%\begin{equation}
%\label{eq:cum_I_gen}
%    \kappa(I_{j_1}, ...,I_{j_l})    = \sum_{i} T(j_1|x_i) T(j_2|x_i)...T(j_r|x_i) \kappa^{(r)}(q_{i}),
%\end{equation}
\begin{multline}
\label{eq:cum_I_gen}
    \kappa(I_{j_1}, ...,I_{j_r})    = \\ = \kappa \left( \sum_{i_1} T(j_1|x_{i_1}) q_{i_1}, ... , \sum_{i_r} T(j_l|x_{i_r}) q_{i_r}\right)  = \\ \overset{\textrm{(I)}}{=}\sum_{i_1,...i_r} T(j_1|x_{i_1}) ... T(j_l| x_{i_r}) \kappa(q_{i_1}, ..., q_{i_r}) =  \\\overset{\textrm{(II)}}{=} \sum_{i} T(j_1|x_i) T(j_2|x_i)...T(j_r|x_i) \kappa^{(r)}(q_{i}),
\end{multline}
 in (I) we used cumulants linearity, in (II) we used the fact that cumulants of independent random variables vanish. We got extension of \eqref{eq:SOFI} for arbitrary measurement and arbitrary multi-detectors correlations; for cumulants, effective transfer functions are products of intensity transfer functions. 

To proceed further, let us assume that the ratio $\tilde \kappa_r = \kappa^{(r)}(q_i) / \langle q_i \rangle$ is the same for all emitters. This assumption is true for statistically identical sources, but also more generally, when each single source consists of multiple independent, identical emitters stick together. Importantly, if blinking dynamics is not known, coefficients $\tilde \kappa_r$  can be estimated based on measured signal, see Sec.~\ref{sec:k_est} for more details. Let $T(j|x) = \sum_\mu A_{\mu, j} (x/\sigma)^\mu$ be Taylor expansions of transfer functions, then \eqref{eq:cum_I_gen} becomes
\begin{multline}
\label{eq:cum_I_ser}
\kappa(I_{j_1}, I_{j_2},...,I_{j_r})    = \tilde \kappa_r \sum_{\mu} A_{\mu, \vec{j}} \theta_\mu, \quad\vec{ j} = (j_1, j_2,...,j_r), \\
A_{\mu, \vec{j}} = \sum_{\mu_1+\mu_2+...+\mu_r = \mu} A_{\mu_1, j_1} A_{\mu_2, j_2}...A_{\mu_r, j_r},
\end{multline}
where $\theta_\mu = \sum _i \langle q_i \rangle (x_i/\sigma)^\mu$ are object's dimensionless spatial moments, previous definition was generalized to fluctuating sources.
%To connect cumulants of intensities $I$ with observable cumulants of photon counts $n$, we prove that (see ...)

%\begin{equation}
   % \kappa \left(n_{j_1}^{(r_1)},...,n_{j_l}^{(r_l)}\right) = \sum_{i_1 = 0}^{r_1}  ... \sum_{i_k=0}^{r_k} \stirling{r_1}{i_1} ... \stirling{r_k}{i_k} \kappa \left( I_1^{(j_1)},...,I_k^{(j_k)} \right),
%\end{equation}

%To see this more directly, let us consider SPADE measurement. Then, in subdiffraction limit
Equation \eqref{eq:cum_I_ser} shows that intensity cumulants provide additional information about the object’s spatial moments. For SPADE measurement, in a subdiffraction limit, \eqref{eq:cum_I_ser} reduces to
\begin{equation}
\label{eq:cum_subdiff_SPADE}
    \kappa(I_{j_1}^\textrm{S}, ...,I_{j_r}^\textrm{S})    = \frac{\tilde \kappa_r}{ \prod_{l=1}^r4^{j_l} j_l!} \theta_{2(j_1+...+j_r)} 
    + \mathcal{O}(\Delta^{2(j_1+...+j_r)+2}),
\end{equation}
so higher-order moments can be inferred from lower-order H-G modes by analyzing suitable cumulants.
For example, $\kappa^{(2)}(I_1) = \frac{\tilde \kappa_2}{16} \theta_4 + \mathcal{O}(\Delta^6)$, so the variance of intensity in 1st H-G mode encodes the fourth spatial moment.
 
 This allows us to enhance estimation precision by combining information from different cumulants: $\theta_4$ can be estimated using both $\kappa(I_2) = \langle I_2 \rangle$ and $\kappa^{(2)}(I_1)$, $\theta_6$ can be estimated using $\kappa(I_3) = \langle I_3 \rangle$, $\kappa(I_1, I_2)$, $\kappa^{(3)}(I_1)$, etc., see Fig.~\ref{fig:FESPADE_principle}c. This extra information sources lead to higher estimation precision, as we will demonstrate in Sec.~\ref{sec:simulation}.

Moreover, one can simplify the measurement and reduce the number of distinguished H-G modes while retaining the ability to estimate higher moments; for example, $\theta_4$ can be estimated \textit{only} based on $\kappa^{(2)}(I_1)$, $\theta_6$ based on $\kappa^{(3)}(I_1)$, etc. Therefore, measurement of just two H-G modes, $\phi_0$ and $\phi_1$,  allows us to estimate all even moments.

Pursuing this idea further, one can use a significantly simpler III measurement instead of SPADE, and still recover all even moments since cumulants of intensity of odd modes satisfy
\begin{equation}
    \kappa^{ (r) \textrm{III} } (I_-) = \frac{\tilde \kappa_r}{4^r} \theta_{2r} + \mathcal{O}(\Delta^{2r+2}),
\end{equation}
so $r$th cumulant encodes $2r$th spatial moment, see Fig.~\ref{fig:FESPADE_principle}d. This is the basis of our newly introduced SOFIII technique.

For odd moments, one needs to use iSPADE, for which cumulants also allow us to gain more information about higher moments. In particular, iSPADE with just four different outcomes---$0$, $0_+$, $0_-$ and $1$---allows us to reconstruct all  moments, for example, using equations
\begin{multline}
\label{eq:cum_bin_iSPADE}
    \kappa^{(r)} (I_1^\textrm{iS}) = \frac{\tilde \kappa_r}{8^r} \theta_{2r} + \mathcal{O}(\Delta^{2r+2}), \\
    \kappa(I_1^{\textrm{iS} (r)},0_+) - \kappa(I_1^{\textrm{iS} (r)},0_-) = \frac{\tilde \kappa_r}{2^{3r+1}} \theta_{2r+1} + \mathcal{O}(\Delta^{2r+3}).
\end{multline}
So far we focused on subdiffraction limit and provided formulas for cumulants in the leading orders of $\Delta$ to underly the most important aspects of fluctuation enhanced SPADE. Nevertheless,  \eqref{eq:cum_I_ser} is valid beyond this approximation and allows us to  express any intensity cumulant as linear combination of spatial moments. This allows us to estimate moments of any object with increased precision, as we will demonstrate in what follows.

\subsection{Estimator construction}
\label{sec:est_construction}
Let us now analyze the statistics of data obtained in a general imaging scheme of fluctuating sources. The observation time is divided into $M$  frames (as was done to describe SOFI in Sec.~\ref{sec:SOFI}), the number of photon counts at detector $j$ in frame $m$ is $n_{j,m}$. To simplify the analysis, we assume that subsequent frames are statistically independent, but we will comment how to go beyond this assumption at the end of this section.

Using collected data, we compute sample temporal moments
\begin{equation}
\label{eq:samp_moments}
    \hat \mu(n_{j_1},...n_{j_r}) = \frac{1}{M} \sum_{m=1}^M n_{j_1,m}...n_{j_r,m},
\end{equation}
which are efficient and unbiased estimators of theoretical temporal moments $ \mu(n_{j_1},...,n_{j_r}) =\langle n_{j_1}...n_{j_r}\rangle$. Then, we construct cumulants estimators using \eqref{eq:mom_to_cum},
\begin{equation}
\label{eq:cumn_est}
    \hat \kappa (n_{j_1}, ...n_{j_r}) = \sum_\pi \Omega_\pi \prod_{B \in \pi} \hat \mu(n_{j_i}:i\in B),
\end{equation}
these estimators are asymptotically unbiased for $M \rightarrow \infty$.

In order to make practical use of \eqref{eq:cum_I_ser}, we should transform estimators of photon count cumulants into estimators of intensity cumulants using
\begin{multline}
    \label{eq:cum_n_cum_I}
    \hat \kappa^{(r_1,...,r_l)} \left(I_{j_1},...,I_{j_l}\right) = \\ = \sum_{k_1 = 0}^{r_1}  ... \sum_{k_l=0}^{r_l} s(r_1,k_1) ... s(r_l,k_l) \hat \kappa^{(k_1,...,k_l)} \left( n_{j_1},...,n_{j_l}\right),
\end{multline}
where $s(r,k)$ denote Stirling numbers of the 1st type, $j_1$,...$j_l$ are non-repeating indices---see Appendix~\ref{app:n_cum_I_cum} for the definition of Stirling numbers and proof of \eqref{eq:cum_n_cum_I}. Formula similar to \eqref{eq:cum_n_cum_I} was derived in Ref.~\cite{Picariello2025} for more general photon statistics---we provide an alternative proof valid for super-poissonian sources only, but extend the applicability to many detectors cases. 

Let vector  $ \hat{ \vec \kappa}^{(I)}$ collect estimators of all intensity cumulants we want to use. For example, with two detectors and cumulants up to 2nd order, $\hat{\vec \kappa}^{(I)} =\left[ \hat{\kappa}(I_1), \hat{\kappa}(I_2), \hat{\kappa}^{(2)}(I_1), \hat{\kappa}^{(2)}(I_2), \hat{\kappa}(I_1,I_2)\right]^T$.

Using frames independce and central limit theorem, we prove (in Appendix~\ref{app:estimator}) that cumulants estimators calculated using  \eqref{eq:cumn_est} and \eqref{eq:cum_n_cum_I} are asymptotically normally distributed, 

\begin{equation}
\label{eq:cum_I_model}
    \hat{\vec \kappa}^{(I)} \overset{M \rightarrow \infty}{\longrightarrow} \mathcal{N}\left(\boldsymbol{A} \vec \theta, \boldsymbol{\Sigma}^{(\kappa)}/M\right),
\end{equation}
where $\boldsymbol{\Sigma}^{(\kappa)}/M$ is a covariance matrix,  $\vec \kappa^{(I)} = \boldsymbol{A} \vec \theta$ is the vector of theoretical cumulants, depending linearly on objects spatial moments collected in  $\vec \theta$. The matrix $\boldsymbol{A}$, whose elements are determined via \eqref{eq:cum_I_ser}, specifies this linear relation. 
Strictly speaking, $\vec \theta $ is usually infinite, since elements of $\vec \kappa^{(I)}$ generally depend on infinitely many spatial moments. In practice, the series is truncated by neglecting higher moments contributions, making $\vec \theta$ finite.

To estimate spatial moments, we use maximum likelihood (ML) estimator, which, for our linear Gaussian model, is given by \cite{kay1993fundamentals}
\begin{equation}
\label{eq:theta_est}
    \hat{\vec \theta} = \left(\boldsymbol{A}^T \left(\hat{\boldsymbol{\Sigma}}^{(\kappa)}\right)^{-1} \boldsymbol{A}\right)^{-1} \boldsymbol{A}^T \left(\hat{\boldsymbol{\Sigma}}^{(\kappa)}\right)^{-1} \hat{\vec{\kappa}}^{(I)},
\end{equation}
where $\hat{\boldsymbol{\Sigma}}^{(\kappa)}$ is a covariance matrix estimated from collected data---the exact procedure of covariance matrix estimation is described in Appendix~\ref{app:estimator}.

To summarize, we provided spatial moments estimation procedure valid for any object size (also beyond subdiffraction limit). One should firstly compute relevant temporal moments estimators directly from collected data using \eqref{eq:samp_moments}, and then transform them to intensity cumulants estimators using \eqref{eq:cumn_est} and \eqref{eq:cum_n_cum_I}. Then, spatial moments estimators are linear combinations of intensity cumulants estimators, with coefficients given by \eqref{eq:theta_est}.

This reasoning can be extended to temporally correlated frames case provided that correlations decay fast enough, so that generalized versions of central limit theorem \cite{Rosenblatt1956} can be used to justify \eqref{eq:cum_I_model}. This can be done, for example, when emitters' blinking is governed by a Markov process, which leads to exponentially decaying temporal correlations between frames \cite{kurdzialek2021super}. In such a case, one needs to take into account these correlations when computing covariance matrix estimator.

\subsection{Fundamental bounds}
When consecutive frames are independent and identically distributed, all relevant information is contained in the empirical temporal moments of photon counts, $\hat{\mu}(n_{j_1},\ldots,n_{j_r})$, provided that a sufficiently high moment order is used.
 However, the question remains whether the procedure described in the previous section optimally estimates object's spatial moments $\theta_\mu$, especially for finite $M$, when  \eqref{eq:cum_I_model} does not hold exactly, ML estimator \eqref{eq:theta_est} is not guaranteed to be optimal, and covariance matrix estimator $\hat \Sigma^{(\kappa)}$ might be inaccurate. 

To answer this question, we will compare the MSE of  the described estimation procedure with the fundamental  C-R bound. Let $\boldsymbol{C}$ be a covariance matrix of spatial moments estimators, $\boldsymbol{C}_{\mu \nu} = \textrm{Cov}(\hat \theta_\mu, \hat \theta_\nu)$.  The C-R bound states that for any unbiased estimators \cite{kay1993fundamentals}
\begin{equation}
\label{eq:multivariate_CR}
    \boldsymbol{C} \ge \boldsymbol{F}^{-1},~~\boldsymbol{F}_{\mu \nu} = \int d \vec x \frac{\partial_\mu p(\vec x | \vec \theta) \partial_\nu p(\vec x|\vec \theta)}{p(\vec x|\vec \theta)}
\end{equation}
where $\boldsymbol{F}$ is a Fisher Information (FI) matrix, $\boldsymbol{A} \ge \boldsymbol{B}$ means that matrix $\boldsymbol{A} - \boldsymbol{B}$ is positive semidefinite, $\vec x$ collects all relevant measurement outcomes, $p(\vec x | \vec \theta)$ is a probability of obtaining outcome $\vec x$ for parameters values $\vec \theta$, $\partial_\mu = \frac{\partial}{\partial \theta_\mu}$. In our case, $\vec{\theta}$ contains the spatial moments to be estimated, and $\vec{x}$ corresponds to the empirical temporal moments of photon counts, $\vec{x}=\hat{\vec{\mu}}^{(n)}$. 
When bounding the precision achievable with a chosen set of cumulants, $\vec{x}$ should include only the temporal moments required to reconstruct those cumulants.

Using \eqref{eq:multivariate_CR}, asymptotic normality of $\hat {\vec \mu}^{(n)} $, and the FI formula for normal distributions\cite{kay1993fundamentals}, we obtain the following bound for MSE of estimating $\theta_\mu$ from $M$ frames:
\begin{equation}
\label{eq:CR_gaussian}
    \delta^2 \hat \theta_\mu \ge \frac{1}{M} \left[ \boldsymbol{F}^{-1} \right]_{\mu \mu},~~ \boldsymbol{F}_{\mu \nu} = \frac{\partial \vec \mu^{(n) T}}{\partial \theta_\mu} {\boldsymbol\Sigma^{(n)}}^{-1} \frac{\partial \vec \mu^{(n)}}{\partial \theta_\nu},
\end{equation}
where $\vec \mu$ collects the \textit{theoretical} temporal moments (the expected values of photon counts products), $\boldsymbol \Sigma^{(n)}$ is a covariance matrix of these photon counts products, see Appendix~\ref{app:estimator} for more details.
Importantly, $\mathbf{F}$ does not depend on $M$ and can be computed exactly from the statistical model of a single frame, without random sampling.
While the bound above is valid for any $M$, it becomes tight only in the asymptotic limit $M \to \infty$. 
Nevertheless, we show that even for finite $M$, the simulated MSEs of our estimators closely follow the bound---this demonstrates both the bound relevance and the estimator optimality.
\section{Experiment simulation}
\label{sec:simulation}
\begin{figure}
    \centering
    \includegraphics[width = 1.0\columnwidth]{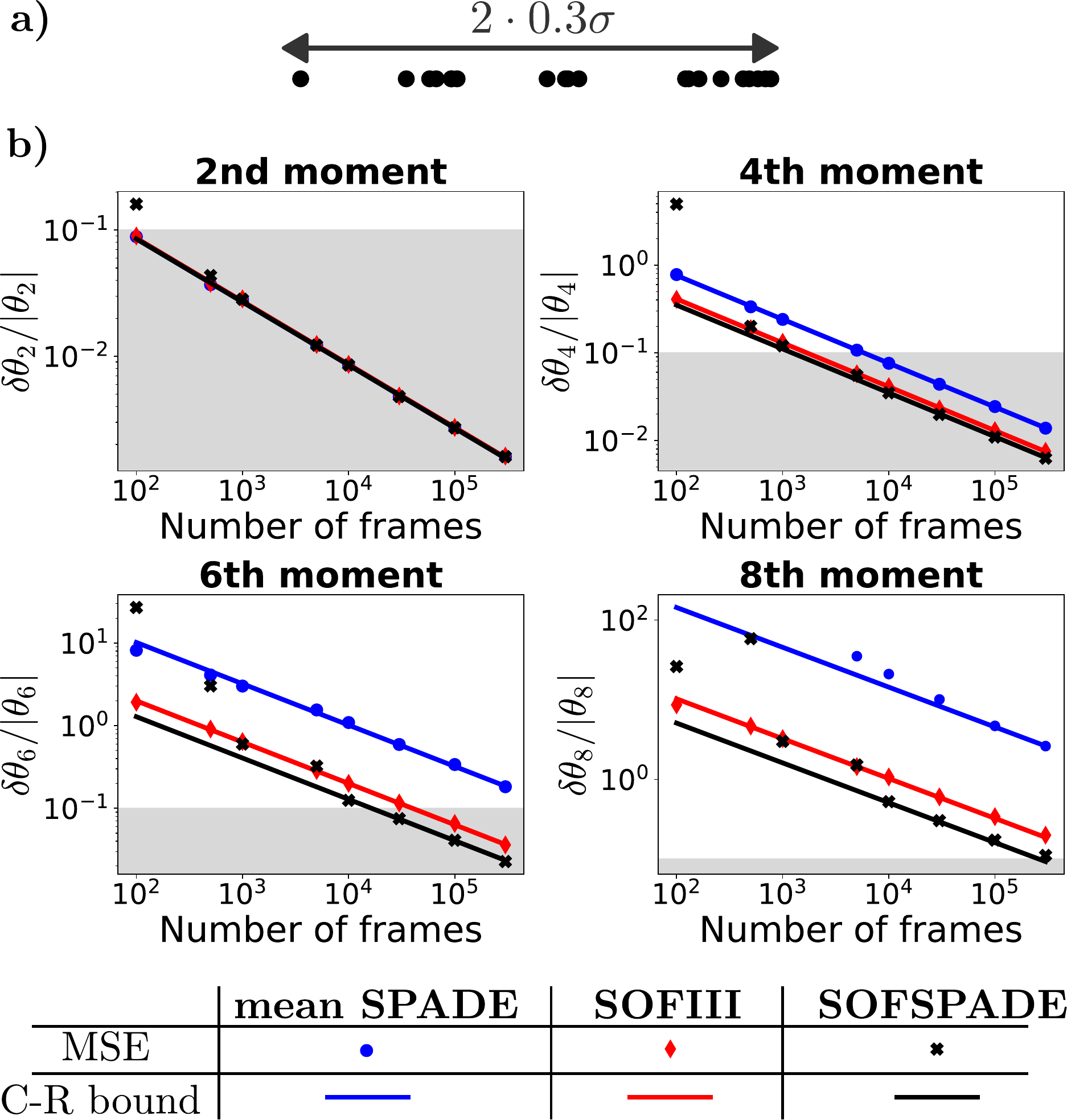}
    \caption{Simulations are performed for a subdiffraction object ($\Delta = 0.3$) composed of 
$k=20$ emitters distributed as in (a).
Panel (b) shows the relative errors of even spatial-moment 
estimates as functions of the 
number of frames  (markers: simulated MSEs; solid lines: C-R bounds)  The area corresponding to accuracy 10\% or better is shaded in each chart to facilitate comparison. Higher-order moments are harder to estimate, and the benefit of fluctuation-based methods (SOFSPADE, SOFIII) 
increases with moment order.}
    \label{fig:SPADE_plots}
\end{figure}
We now demonstrate the performance of the proposed imaging techniques, SOFSPADE and SOFIII, in estimating spatial moments of blinking objects.  
We do not discuss the subsequent reconstruction of the object from its moments, as this has already been demonstrated in SPADE imaging simulations~\cite{Bearne2021} and experiments~\cite{Pushkina2021, Frank2023Optica, Duplinskiy2025}, both in confocal and wide-field setups.  
It is clear that improved accuracy in estimating spatial moments—especially higher-order ones—directly translates into higher image resolution and reconstruction fidelity.

Consider an object consisting of $k=20$  independent blinking point emitters, distributed within a subdiffraction region ($\Delta = 0.3$), as shown in Fig.~\ref{fig:SPADE_plots}a. Each emitter switches between ``on'' and ``off'' states with brightnesses $q_\mathrm{on}=100\frac{\textrm{photons}}{\textrm{frame}}$ and $q_\mathrm{off}=5\frac{\textrm{photons}}{\textrm{frame}}$, respectively, the probability of ``on'' state is $p_\mathrm{on}=0.1$---these are reasonable assumptions for SOFI experiments \cite{Dedecker2012, Deschout2016, Vandenberg2017}.  For each frame, the brightness $q_i$ of every emitter is drawn independently, after which photon counts at the detectors are sampled from a Poisson distribution with mean given by~\eqref{eq:poiss_gen}.  

This process is repeated independently for $M$ frames. From resulting time series of photon counts, empirical temporal moments are computed using~\eqref{eq:samp_moments}, and the corresponding spatial-moment estimators $\hat{\theta}_\mu$ are obtained via~\eqref{eq:cumn_est}, \eqref{eq:cum_n_cum_I}, \eqref{eq:theta_est}.  
The entire estimation procedure is repeated $N=1000$ times for independently generated datasets of $M$ frames, and for each run we evaluate the squared error $(\hat{\theta}_\mu - \theta_\mu)^2$.  
Averaging these errors over all repetitions yields the MSE for each estimated moment. We compare these MSEs with C-R bounds obtained using  \eqref{eq:CR_gaussian}. 

%\begin{figure*}[t]
%\includegraphics[width = 0.95 \linewidth]{figs/SPADE_horizontal.pdf}
%\caption{a}
%\label{fig:SPADE}
%\end{figure*}

Let us firstly consider estimation of 5 leading even moments, $\vec \theta = [\theta_0, \theta_2, \theta_4, \theta_6, \theta_8]^T$. This can be done using mean intensities on 5 SPADE detectors (mean SPADE), which corresponds to $\hat {\vec \kappa} = [ \hat \kappa(I_0^\textrm{S}),\hat \kappa(I_1^\textrm{S}),\hat \kappa(I_2^\textrm{S}),\hat \kappa(I_3^\textrm{S}),\hat \kappa(I_4^\textrm{S})]^T$. Alternatively, one can significantly simplify the measurement and use SOFIII technique, which requires just two detectors and temporal cumulants up to 4th order, $\hat {\vec \kappa} = [\hat{\kappa}(I_+^\textrm{III}), \hat{\kappa}(I_-^\textrm{III}), \hat{\kappa}^{(2)}(I_-^\textrm{III}), \hat{\kappa}^{(3)}(I_-^\textrm{III}), \hat{\kappa}^{(4)}(I_-^\textrm{III})] $. Finally, to get maximal precision, one can use SOFSPADE, in which temporal correlations between multiple detectors are used,
\begin{multline*}
\hat {\vec \kappa} = [ \hat \kappa(I_0^\textrm{S}),\hat \kappa(I_1^\textrm{S}),\hat \kappa(I_2^\textrm{S}),\hat \kappa(I_3^\textrm{S}), \hat \kappa(I_4^\textrm{S}),\hat \kappa^{(2)}(I_1^\textrm{S}),\hat \kappa(I_1^\textrm{S}, I_2^\textrm{S}), \\ \hat \kappa(I_1^\textrm{S}, I_3^\textrm{S}), \hat \kappa^{(2)}(I_2^\textrm{S}),\hat \kappa^{(3)}(I_1^\textrm{S}), \hat \kappa^{(2,1)}(I_1^\textrm{S}, I_2^\textrm{S}),\hat \kappa^{(4)}(I_1^\textrm{S})]^T ,
\end{multline*}
note that we neglected cumulants with sum of detectors labels greater than 4 since they do not give any information about spatial moments of orders up to $8$.
%\begin{figure*}[t]
%\includegraphics[width = 0.95 \linewidth]{figs/iSPADE_horizontal2.pdf}
%\caption{a}
%\label{fig:iSPADE}
%\end{figure*}
The calculated  MSEs and corresponding C-R bounds as functions of the number of frames $M$ are shown in Fig.~\ref{fig:SPADE_plots}b. These results demonstrate that higher-order spatial moments can be reliably estimated using the experimentally accessible SOFIII technique.  Remarkably, SOFIII achieves even higher precision than full SPADE with mean signal only. This advantage diminishes for weaker fluctuations, where SOFIII may become less precise than mean SPADE, though still far easier to implement. However,  SOFSPADE consistently outperforms both methods, as it combines the information from multiple detectors and temporal correlations.  Our simulations show that SOFSPADE advantage grows with spatial moment order  since higher spatial moments receive contributions from multiple cumulants. For $\theta_2$, SOFSPADE gives almost no advantage, as only the first-order cumulant $\langle I_1 \rangle$ carries information about this parameter. In contrast, higher-order cumulants dramatically enhance higher-order moments estimates: SOFSPADE achieves the same precision for the 8th moment with nearly $1000$ times fewer frames than mean-signal SPADE. 

\begin{figure}
    \centering
    \includegraphics[width = 1.0\columnwidth]{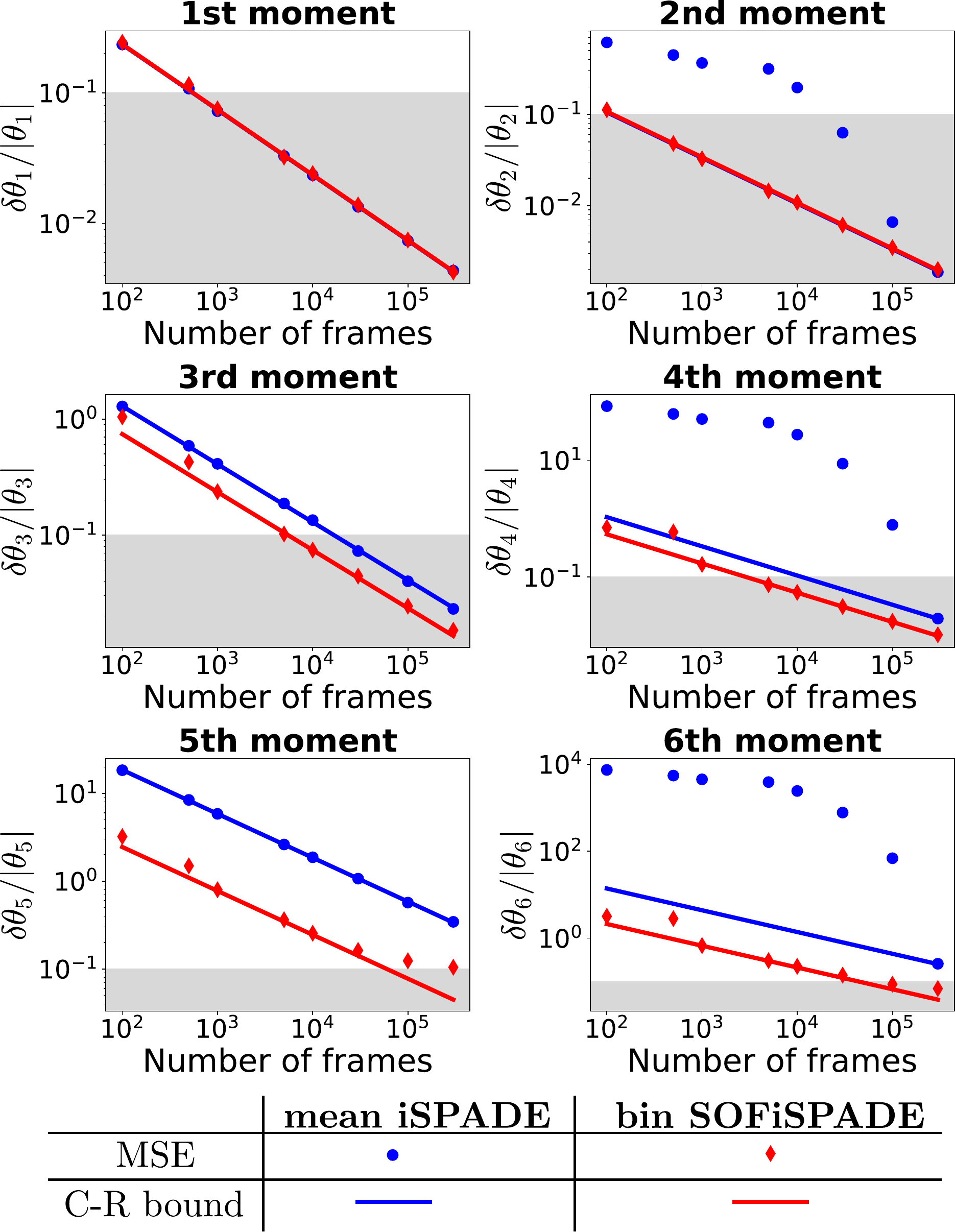}
    \caption{Odd and even moments of an object shown in Fig.~\ref{fig:SPADE_plots}a are estimated using mean iSPADE method or bin SOFiSPADE, where measurement is simplified, and temporal fluctuations are used. The simulated MSEs and corresponding C-R bounds of both methods are plotted as functions of the number of frames. For even moments, the estimator based on mean iSPADE only becomes efficient for large number of frames, but bin SOFiSPADE method is free of this issue. }
    \label{fig:iSPADE_plots}
\end{figure}

To estimate odd moments, we need to use iSPADE measurement. If we want to know moments up to order 6, then  $\vec \theta = [\theta_0,\theta_1,...,\theta_6]$. Let us consider two estimation strategies: (i) mean iSPADE, for which we use mean intensities on 8 detectors, $\hat{\vec \kappa} = [\kappa(I_{0}^\textrm{iS}), \kappa(I_{0_\pm}^\textrm{iS}), \kappa(I_{1_\pm}^\textrm{iS}), \kappa(I_{2_\pm}^\textrm{iS}),\kappa(I_{3}^\textrm{iS})]^T$; (ii) binary stochastic optical fluctuation iSPADE (bin SOFiSPADE), for which 2 H-G modes are interfered, which leads to 4 detectors $0$, $0_+$, $0_-$, $1$, for which temporal cumulants from \eqref{eq:cum_bin_iSPADE} are evaluated, $\hat {\vec \kappa} = [\hat \kappa(I_0), \{ \hat \kappa^{(r)}(I_1), \hat \kappa^{(1,r-1)}(I_{0_\pm }, I_1 ) : r \in \{1,2,3\}\} ]^T$.
The estimation MSEs again follow the C-R bound for sufficiently large $M$, see Fig.~\ref{fig:iSPADE_plots}. Interestingly, fluctuation-based method with simplified measurement again outperforms mean iSPADE, especially for high even moments and small values of $M$, for which C-R bound for mean iSPADE is not saturated. 

Moments of order higher than 8 are close to 0 for the considered object, so their absence in the estimation procedure is well-justified. However, for larger objects, it might be necessary to estimate higher-order moments to avoid systematic estimation errors coming from implicit assumption that these moments are equal to 0, made when neglecting higher-moment terms in \eqref{eq:cum_I_ser}. See Appendix~\ref{app:obj_size} for further discussion of this issue and simulation results for moments estimation of objects of different sizes $\Delta$.

\subsection{The role of dark counts}
\label{sec:dark_counts}
So far, we have considered an idealized scenario, neglecting all noise sources except the fundamentally unavoidable shot noise. However, realistic detectors also suffer from additional noise sources, such as dark counts, which are known to significantly affect the performance of SPADE-based imaging \cite{Len2020, Lupo2020, Oh2021}. This is because much of the information about sub-diffraction objects comes from the almost-dark ports corresponding to higher-order H-G modes. Unfortunately, this dark-port information is highly sensitive to additional noise: once the number of dark counts becomes comparable to or larger than the expected signal, the information is almost entirely lost.

Fortunately, temporal cumulants of the lower-order H-G modes are much less affected by dark counts, which makes SOFSPADE and SOFIII substantially more robust to this type of noise. An analogous effect is observed when upgrading DI to SOFI \cite{Dertinger2009}, but because dark counts are especially detrimental to SPADE, the robustness gained from exploiting fluctuations is even more critical here.

\begin{figure}[h]
    \centering
    \includegraphics[width = 1.0\columnwidth]{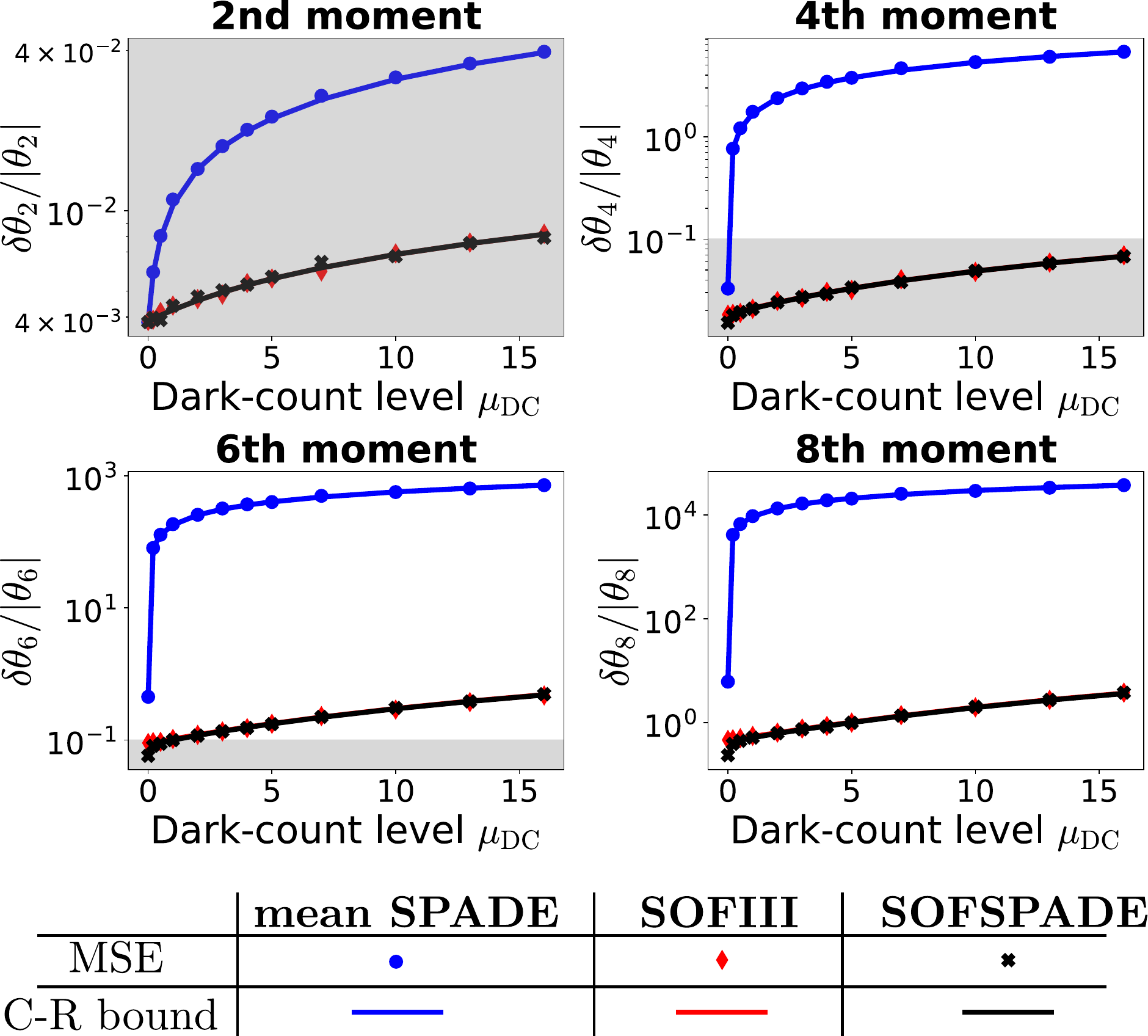}
    \caption{Even moments of an object shown in Fig.~\ref{fig:SPADE_plots}a are estimated in the presence of dark-count noise, with a mean number of dark counts per detector per frame equal to $\mu_\textrm{DC}$. The simulated MSEs and corresponding C-R bounds  of mean SPADE , SOFSPADE  and SOFIII  methods are plotted as functions of $\mu_\textrm{DC}$ for a fixed number of frames $M=50000$.
 SOFIII and SOFSPADE  perform equally well and significantly outperform mean SPADE in the presence of dark counts. }
    \label{fig:SPADE_plots_DC}
\end{figure}

Let us assume that the numbers of dark counts recorded by each detector in each frame are independent Poissonian random variables with mean $\mu_\textrm{DC}$. We numerically simulate the reconstruction  of even moments for the object depicted in Fig.~\ref{fig:SPADE_plots}a for a fixed number of frames $M=50000$ and different levels of dark-count noise $\mu_\textrm{DC}$; the emitters' blinking dynamics is assumed to be the same as in the previous simulations. Both the estimator construction and the C-R bound formula are adjusted to properly account for this additional noise---see Appendix~\ref{app:estimator} for details. The  estimation MSEs and C-R bounds as a function of $\mu_\textrm{DC}$ for mean SPADE, SOFSPADE and SOFIII methods are shown in Fig.~\ref{fig:SPADE_plots_DC}.

The advantage of SOFSPADE/SOFIII over mean SPADE grows drastically with dark-count noise. For $\mu_\textrm{DC} \gtrsim 1$, the relative error for the 8th moment is $\sim 10^4$ times smaller than for mean SPADE---equivalent to a $10^8$-fold increase in frame number if temporal-cumulant information were unavailable.

SOFIII and SOFSPADE perform equally well across all estimated moments in the presence of dark counts, indicating that the information carried by the first H-G mode's temporal cumulants, $\kappa^{(r)}(I_1)$, dominates over that from higher-order modes.

Notably, SOFSPADE/SOFIII also slightly outperform mean SPADE for the 2nd moment---unlike in the dark-count-free case. This seems to contradict the intuition that 2nd-moment information resides solely in the mean of the first H-G mode (Fig.~\ref{fig:FESPADE_principle}), but that picture holds strictly only as $\Delta \rightarrow 0$. For finite $\Delta$, moment estimators become correlated through off-diagonal FI-matrix terms. As a result, in the presence of dark-count noise, the significantly reduced precision of higher-order moment estimates propagates into the 2nd-moment estimate as well; incorporating temporal-cumulant information counteracts this effect, limiting how much the imprecision of higher moments degrades the 2nd-moment estimate.

%\begin{figure}[h]
%\includegraphics[width = 0.95\columnwidth]{figs/iSPADE_vertical.pdf}
%\end{figure}
%Let us also discuss how fluctuations affect...
\section{Practical aspects}
\label{sec:practical}
\subsection{Experimental implementation}
\label{sec:implementation}

SOFSPADE and SOFIII do not require any optical hardware beyond that of the techniques they build on---SPADE and III, respectively. SOFSPADE uses the same mode-sorting apparatus as conventional SPADE---a multi-plane light converter, or a heterodyne setup with structured local oscillators \cite{Tan2023Optica, Rouviere2024, Santamaria2024, Yang2016, Pushkina2021, Frank2023Optica, Duplinskiy2025}---while SOFIII uses the same interferometer as conventional III \cite{Smith2005, Sandeau2006, Nair2016}. In fact, fluctuations reduce the required hardware complexity compared to their non-fluctuation counterparts, because higher-order moments can be recovered from cumulants of the lower H-G modes---meaning that fewer modes need to be sorted than the number of moments estimated, unlike for mean SPADE. Similarly, estimating odd moments via bin SOFiSPADE needs an interferometric mixing stage between only the two lowest-order modes, rather than the full set of mode pairs required by conventional iSPADE. 

The main additional requirement common to all fluctuation-enhanced techniques discussed here is time-resolved detection: photon counts must be recorded frame by frame so that temporal cumulants can be computed. This is standard practice in SOFI-based fluorescence microscopy \cite{Dertinger2009, Dertinger2010} and does not require specialized hardware beyond a sufficiently fast camera (or sufficiently fast bucket detectors at the output of SPADE/III device).  The emitters themselves must  exhibit intensity fluctuations---such fluctuations are observed in quantum dots \cite{efros2016origin} or organic dyes \cite{dickson1997off}, commonly used as labels in fluorescence microscopy. The camera acquisition time must be chosen such that emitters' fluctuations are not averaged out within a single frame. 

\subsection{Generalization to 2D objects and non-Gaussian PSFs}
\label{sec:2D}
To illustrate the working principle of SOFSPADE and SOFIII superresolution techniques without introducing extra mathematical complications, we focused on 1D  imaging. Let us now discuss how these results generalize to a more practical, 2D case. The SPADE measurement is then performed in the basis of 2D H-G modes defined as  $\phi_{jk}(x',y') = \phi_j(x') \phi_k(y')$, $j,k \in \{0,1,2,...\}$. The corresponding transfer function is a product of 1D transfer functions,
\begin{equation}
    T^\textrm{2S}(j,k|x,y) = T^S(j|x) T^S(k|y),
\end{equation}
where $x,y$ are coordinates of  a light emitter in the object plane. A 2D object consisting of point sources with positions $(x_i, y_i)$ and fluctuating brightnesses $q_i$ can be characterized by its 2D moments defined as $\theta_{\mu_x, \mu_y} = \sum_i \langle q_i \rangle x_i^{\mu_x} y_i^{\mu_y}$. As in 1D case, objects moments are directly related to  mean intensities $I_{j,k}$  of different H-G modes---for subdiffraction objects
\begin{equation}
    \langle I_{j,k}^\textrm{2S} \rangle = \frac{1}{4^{j+k}j! k!} \theta_{2j, 2k} + \mathcal{O}(\Delta^{2j+2k+2}),
\end{equation}
where $\Delta = \max_i \sqrt{x_i^2 + y_i^2}/\sigma$ characterizes the size of 2D object. Again, one can go beyond subdiffraction limit and express $\langle I_{j,k} \rangle$ as a combination of infinite number of moments with leading contribution from $\theta_{2j,2k}$, as in \eqref{eq:SPADE_Ij}. 

The 2D analogue of \eqref{eq:cum_subdiff_SPADE} is
\begin{multline}
    \kappa(I_{j_1, k_1}^\textrm{2S},..., I_{j_r, k_r}^\textrm{2S}) = \\ = \frac {\tilde \kappa_r}{\prod_{l=1}^r 4^{j_l+k_l} j_l!k_l!} \theta_{2(j_1+...+j_r), 2(k_1+...+k_r)} , 
\end{multline}
so as in 1D case, cumulants of intensities in lower modes provide information about higher-order moments; for example, $\kappa(I_{0,1}^\textrm{2S}, I_{1,0}^\textrm{2S}) = \frac{\tilde \kappa_2}{16} \theta_{2,2} + \mathcal{O}(\Delta^6)$ $\kappa^{(2)}(I^\textrm{2S}_{1,1}) = \frac{\tilde \kappa_2}{256} \theta_{4,4} +\mathcal{O}(\Delta^{10})$, etc. Therefore, fluctuations help to estimate even 2D moments $\theta_{j,k}$ for which $j+k \ge 4$; the general estimator construction is the same as for 1D case. To estimate odd moments, one should use 2D version of iSPADE measurement \cite{Tsang_2017}; there is also 2D version of III technique \cite{Zheng2017}, which, when enhanced by fluctuations, may be used to estimate all 2D even moments.

For non-Gaussian PSFs, the measurement modes $\phi_j$ (or $\phi_{jk}$ in two dimensions) should be adapted to the specific PSF. A general construction of such modes for an arbitrary PSF was introduced in Ref.~\cite{Rehacek2017} in the context of two-point separation estimation and was subsequently extended to spatial-moment estimation in Ref.~\cite{Tsang_2017}. The corresponding transfer function, $T(j|x)$ or $T(j,k|x,y)$, is obtained from the squared overlap between the measurement mode and the shifted coherent PSF, i.e., the optical field produced by a single point source. Expanding this transfer function in a Taylor series then yields the coefficients $A_{\mu,j}$ appearing in Eq.~\eqref{eq:SPADE_Ij} for the modified PSF and measurement basis. These coefficients were calculated explicitly for a circular aperture in Ref.~\cite{Tsang_2017}. Once the corresponding coefficients $A_{\mu,j}$ have been determined, the temporal cumulant-based estimation procedure developed in this work can be applied without further modification.
%\subsection{Non-Gaussian PSFs}
%In Ref.~\cite{Rehacek2017} it was shown how to construct optimal measurement basis modes $\phi_j$ for two-point separation estimation assuming an arbitrary shape of the PSF $U$. The result was then extended to the moments estimation problem in Ref.~\cite{Tsang_2017}
\subsection{Blinking parameters estimation}
\label{sec:k_est}
To apply SOFSPADE or SOFIII techniques in practice, one needs to know $\tilde \kappa_r$ coefficients characterizing blinking dynamics of emitters. These coefficients can be easily computed when blinking parameters $q_\textrm{on}$, $q_\textrm{off}$ and $p_\textrm{on}$ are known. If this is not true, one can estimate $\tilde \kappa_r$ from experimental data. Let $n_{\textrm{all},m} = \sum_j n_{j,m}$ be the total number of photon counts collected from all detectors in a given frame $m$. Then, one can calculate sample moments $\hat \mu^{(r)} (n_{\textrm{all}}) = \frac{1}{M} \sum_{m=1}^M n^r_{\textrm{all},m}$, and, further on, intensity cumulants of total signal $\hat \kappa^{(r)} (I_{\textrm{all}})$ using \eqref{eq:cumn_est} and \eqref{eq:cum_n_cum_I}. Then, coefficients $\tilde \kappa_r$ can be estimated as $\tilde \kappa_r = \hat \kappa^{(r)} (I_{\textrm{all}}) / \langle I_\textrm{all} \rangle $---because of cumulants additivity, the ratio between $r$th cumulant and mean of total signal is the same as the corresponding ratio for a single emitter.

Realistic fluorophores are subject to photobleaching, which gradually decreases their mean intensity but, more importantly, may also affect their higher-order temporal statistics---and it is this latter effect that causes the coefficients $\tilde \kappa_r$ to vary in time. 
In this case, the total measurement time should be divided into shorter intervals over which $\tilde \kappa_r$ remain approximately constant, with the coefficients estimated independently within each one. 
The spatial-moment estimator is then built by computing intensity cumulants separately for each interval---so that $\hat{\vec{\kappa}}^{(I)}$ contains the cumulants of all intervals as separate elements---and constructing a matrix  $\boldsymbol{A}$ that accounts for the interval-dependent values of $\tilde \kappa_r$, before applying \eqref{eq:theta_est} to the segmented data. 
Since cumulants from different intervals are independent, the resulting covariance matrix $\boldsymbol{\Sigma}^{(\kappa)}$ is block-diagonal, so the resulting reconstruction complexity only grows linearly with the number of intervals.
\subsection{Cross-talk noise}

Beyond dark-count noise, thoroughly discussed in Sec.~\ref{sec:dark_counts}, SPADE-based techniques may also suffer from cross-talk between modes \cite{Gessner2020}. Spatial moments can still be estimated in this case by modifying the transfer function $T(j|x)$ fed to the reconstruction algorithm to account for the cross-talk, though the accuracy of higher-order moments may degrade---for mean SPADE as well as for SOFSPADE and SOFIII---since leakage from lower-order modes can then dominate the signal collected in higher-order H-G modes. Note, however, that cross-talk reduction is likely easiest for SOFIII, since its underlying field measurement is comparatively simple, requiring only the cross-talk between the two relevant outputs to be considered.

\subsection{Coherence between emitters}

In this work, we assume that all emitters are mutually incoherent, which is a good approximation for most microscopy setups. Partially coherent sources would require a more general treatment, since even \eqref{eq:poiss_gen} no longer holds in this case---one would instead need to know the \textit{coherent} transfer function of the system, which describes how the \textit{complex field amplitude}, rather than light intensity, is transferred from the object to the detector. This would allow one to correctly calculate the measuring device's response to any field configuration of the object. In this case, the cumulants of the detected intensities can no longer, in general, be expressed as a linear combination of spatial moments---instead, one should estimate coherent analogs of spatial moments, which also carry information about the spatial coherence of the object plane.

A physically relevant case in which our framework fails, and needs to be replaced with a coherent one, is the imaging of strong thermal sources\cite{Nair2016PRL, Lupo2016}. There, to obtain the correct photon-count statistics at the detector, one should first consider the coherent field produced in the object plane for fixed emitter amplitudes, then propagate this field to the measuring device, and only afterwards average the result over the different emitter amplitudes.

\section{Conclusions}
\label{sec:discussion}
To summarize, we have shown that blinking can substantially enhance the performance of SPADE: the proposed \textsc{SOFSPADE} method exploits temporal fluctuations to achieve markedly improved precision, especially for higher moments estimation.

More importantly, emitter fluctuations enable the far simpler III measurement to recover the full object information. The proposed \textsc{SOFIII} technique is experimentally much easier to implement than SPADE, yet it still allows recovery of all even spatial moments of the object. The only practical limitation comes from the fact that higher-order cumulants become hard to estimate.

The pronounced robustness of SOFSPADE and SOFIII to dark-count noise, compared to mean SPADE, is a significant practical advantage in realistic, noisy imaging environments. 

It is especially promising to implement SOFIII in a confocal microscope---on one hand, confocal versions of SPADE have already demonstrated the ability to reconstruct complex objects with enhanced resolution; on the other hand, \textsc{SOFIII} delivers the same information content as SPADE, while the III measurement has already been realized for fluorescent sources and high-NA objectives. Incorporating blinking fluorophores and time-resolved detectors should not significantly increase the experimental complexity of an III setup. 

In this work, we analyzed the spatial moments estimation of complex objects. As shown in Appendix~\ref{app:theory}, fluctuations do not enhance SPADE measurement for the simpler task of estimating the separation between two equally bright point emitters. However, as demonstrated in Refs.~\cite{Horoshko2025,Katamadze2025}, non-Poissonian statistics may be beneficial for two emitters imaging, when their centroid is unknown, especially for emitters of unequal brightness.

\emph{Acknowledgements} I would like to thank Rafał Demkowicz-Dobrzański for suggestions, discussions, and reading the initial version of my manuscript. I would also like to thank Konrad Banaszek for his support, and Wojciech Górecki for discussions. This work is a part of the project ''Quantum Optical Technologies''  carried out within the International Research Agendas programme of the Foundation for Polish Science cofinanced by the European Union under the European Regional Development Fund and was supported by the National Science Center (Poland) grant No.2020/37/B/ST2/02134. The author is a recipient of the Foundation for Polish Science START 2025 scholarship. 

\bibliography{biblio}

\appendix
\onecolumngrid
\section{Cumulants properties}
\label{app:cumulants}
We can get explicit formulas for single-variable and multi-variable cumulants by expanding cumulants generating functions, \eqref{eq:cum_single} and \eqref{eq:cum_multi} from the main text respectively. This allows us to express cumulants in terms of moments, see \eqref{eq:mom_to_cum}. The expressions for single-variable cumulants up to order $4$ are:
\begin{subequations}
\label{eq:app_cum_single_explicit}
\begin{align}
\kappa^{(1)}(X) &= \langle X \rangle , \\
\kappa^{(2)}(X) &= \langle X^2 \rangle - \langle X \rangle^2 , \\
\kappa^{(3)}(X) &= \langle X^3 \rangle - 3 \langle X^2 \rangle \langle X \rangle
           + 2 \langle X \rangle^3 , \\
\kappa^{(4)}(X) &= \langle X^4 \rangle - 4 \langle X^3 \rangle \langle X \rangle
           - 3 \langle X^2 \rangle^2
           + 12 \langle X^2 \rangle \langle X \rangle^2
           - 6 \langle X \rangle^4 .
\end{align}
\end{subequations}
Note that 2nd and 3rd cumulants are equivalent to central moments. The analogous expression for multivariable cumulants are
\begin{subequations}
\label{eq:app_cum_multi_explicit}
\begin{align}
\kappa(X,Y) &= \langle XY \rangle - \langle X \rangle \langle Y \rangle , \\[0.5em]
\kappa(X,Y,Z) &= \langle XYZ \rangle
 - \langle XY \rangle \langle Z \rangle
 - \langle XZ \rangle \langle Y \rangle
 - \langle YZ \rangle \langle X \rangle
 + 2 \langle X \rangle \langle Y \rangle \langle Z \rangle , \\[0.5em]
\kappa(X,Y,Z,W) &= \langle XYZW \rangle
 - \langle XYZ \rangle \langle W \rangle
 - \langle XYW \rangle \langle Z \rangle
 - \langle XZW \rangle \langle Y \rangle
 - \langle YZW \rangle \langle X \rangle \nonumber\\
&\quad
 - \langle XY \rangle \langle ZW \rangle
 - \langle XZ \rangle \langle YW \rangle
 - \langle XW \rangle \langle YZ \rangle \nonumber\\
&\quad
 + 2 \big(
   \langle XY \rangle \langle Z \rangle \langle W \rangle
 + \langle XZ \rangle \langle Y \rangle \langle W \rangle
 + \langle XW \rangle \langle Y \rangle \langle Z \rangle \nonumber\\
&\qquad\quad
 + \langle YZ \rangle \langle X \rangle \langle W \rangle
 + \langle YW \rangle \langle X \rangle \langle Z \rangle
 + \langle ZW \rangle \langle X \rangle \langle Y \rangle
 \big) \nonumber\\
&\quad
 - 6 \langle X \rangle \langle Y \rangle \langle Z \rangle \langle W \rangle .
\end{align}
\end{subequations}
One can also express moments in terms of cumulants; inverted relation \eqref{eq:mom_to_cum} reads
\begin{equation}
\label{eq:app_cum_to_mom}
   \mu(X_1,...,X_r) = \left\langle X_1 ... X_r \right\rangle = \sum_\pi \prod_{B \in \pi} \kappa(X_i:i \in B),
\end{equation}
$\pi$ is set of all partitions of $\{1,...,r\}$ into subsets.
Importantly, multivariate cumulants are linear in each argument
\begin{equation}
\label{eq:app_cum_linearity}
    \kappa(\alpha X + \beta Y, X_2,...,X_n) = \alpha \kappa( X,X_2,...,X_n) + \beta(Y,X_2,...,X_n),
\end{equation}
$\alpha$, $\beta$ are numbers.
For single-variable cumulant of order $r$ this formula leads to
\begin{equation}
\label{eq:app_cum_sum}
    \kappa^{(r)}(\alpha X + \beta Y) =  \sum_{j=0}^r \binom{r}{j} \alpha^j \beta^{r-j} \kappa^{(j,r-j)}(X,Y)
\end{equation}
The crucial property of cumulants , which distinguishes them from other statistics like moments and central moments, is 
\begin{equation}
\label{eq:app_cum_independent}
    \kappa^{(r_1,...,r_n)}(X_1,...,X_n) = 0~~\textrm{for independent}~~X_1,...,X_n~~\textrm{if}~~ \# \{i: r_i >0 \} \ge 2,
\end{equation}
where $\#\{i: \bullet\}$ is the number of indices $i$ satisfying condition $\bullet$.
This property makes cumulants so useful in SOFI, SOFSPADE and SOFIII techniques as long as different light emitters are statistically independent. Using \eqref{eq:app_cum_sum} and \eqref{eq:app_cum_independent}, we get
\begin{equation}
    \kappa^{(r)}(\alpha X + \beta Y) = \alpha^r \kappa^{(r)}(X) + \beta^r \kappa^{(r)}(Y)~~\textrm{for independent}~~X,Y,
\end{equation}
and, more generally
\begin{equation}
    \kappa^{(r)}\left( \sum_i \alpha_i X_i \right) = \sum_i \alpha_i^r~ \kappa^{(r)}(X_i)~\textrm{for independent}~~X_1,...,X_n.
\end{equation}
This equation is a basic principle of SOFI technique, as it directly translates to \eqref{eq:SOFI} when $X_i$ are replaced with $q_i$ and $\alpha_i$ are replace with $U(x_j-x_i)$.  

The other property, which will be used in Sec.~\ref{app:n_cum_I_cum} is the law of total cumulance \cite{Brillinger1969}: when $X_1, X_2,...,X_n,Y$ are random variables, then
\begin{equation}
\label{eq:app_total_cumulance}
    \kappa(X_1,X_2,...,X_n) = \sum_\pi \kappa_Y (\kappa(X_i:i \in B|Y) : B \in \pi),
\end{equation}
where summation is over all partitions $\pi$ of set of indices$\{ 1,2,...,n\}$ into subsets, $B$ denotes a subset of a given partition; for example, when $n=6$, then one exemplary partition of the set of indices is $\pi =\{ \{1,2\},\{3,4\},\{5\},\{6\}\}$, elements of this partition, are $B=\{1,2\}$, $B=\{3,4\}$, $B=\{5\}$, $B=\{6\}$. Conditional cumulants $\kappa(X_i:i \in B|Y)$ depend on the value of $Y$, so they are  random variables because $Y$ is a random variable; we use notation $\kappa_Y$ to indicate that the external cumulant is computed taking into account the randomness of $Y$ only. To get some more intuition, let us see how \eqref{eq:app_total_cumulance} looks like for $n=2$ and $n=3$
\begin{equation}
    \kappa(X_1,X_2) = \kappa_Y(\kappa(X_1,X_2|Y)) + \kappa_Y(\kappa(X_1|Y), \kappa(X_2|Y))
\end{equation}
\begin{multline}
    \kappa(X_1,X_2,X_3) = \kappa_Y(\kappa(X_1,X_2,X_3|Y)) + \kappa_Y(\kappa(X_1|Y), \kappa(X_2,X_3|Y)) + 
\kappa_Y(\kappa(X_2|Y), \kappa(X_1,X_3|Y)) + \\ +
\kappa_Y(\kappa(X_3|Y), \kappa(X_1,X_2|Y)) +
\kappa_Y(\kappa(X_1|Y), \kappa(X_2|Y), \kappa(X_3|Y))
\end{multline}

Finally, another useful property is that all cumulants of a Poissonian random variable are equal to mean value of this variable,
\begin{equation}
\label{eq:app_cum_poiss}
   X \sim \textrm{Poiss}(\mu) \Rightarrow \forall_r~~\kappa^{(r)}(X) = \mu,
\end{equation}

\section{Relation between $n$-cumulants and $I$-cumulants}
\label{app:n_cum_I_cum}
In this work we distinguish between photon-count numbers $n$ and intensities $I$ which are linked by relation $n \sim \textrm{Poiss}(I)$, see \eqref{eq:poiss_gen}. It is much easier to use intensities in derivations because relation $I_j = \sum_i T(j|\boldsymbol{r}_i) q_i$ holds for $I$ (expected values of Poissonian random variables), but not for $n$---this is related to the fact that coherent (Poissonian) state at beam-splitter transforms into independent coherent states, but Fock state transforms into non-trivially correlated states. In fact, for a fixed numbers of photons emitted by different sources $n_i$, photon counts at detectors $n_j$ are correlated and distributed according to multinomial distribution. We can then include fluctuations by introducing a probability of different numbers of photons emitted $P(n_i)$, and write down the statistical model for $n_j$ as a combination of multinomial distributions---this was done for direct imaging measurement in Ref.~\cite{Picariello2025}.

Here we take different approach, valid for super-poissonian sources only---we assume that the number of photons emitted by each source in a single frame is given by $\textrm{Poiss}(q_i)$, where $q_i$ is a random variable describing brightness of a given source. Therefore, the number of emitted photons fluctuates both due to variability of $q_i$ and due to shot noise, which leads to superpoissonian fluctuations. Then, we  use \eqref{eq:poiss_gen} to transform brightnesses $q_i$ into detector intensities $I_j$, and, further on, into detectors' photon counts $n_j$.

For this, we need \eqref{eq:cum_n_cum_I}, which links measurable cumulants of photon counts with easier-to-calculate cumulants of intensities. Let us now prove \eqref{eq:cum_n_cum_I}. 

Firstly, let us write an equation for $\kappa^{(r_1,...,r_l)} \left( n_{j_1},...,n_{j_l}\right)$ (when $j_1$, ..., $j_l$ are non-repeating indices) using the law of total cumulance \eqref{eq:app_total_cumulance} with $Y = (I_{j_1},...,I_{j_l})$, $X_i = n_{j_i}$; then conditional random variables $X_1,...,X_l|Y$ are independent Poissonian variables with mean values $\langle X_k| Y \rangle = I_{j_k}$. Using the fact that cumulants of independent random variables vanish \eqref{eq:app_cum_independent} and using  \eqref{eq:app_cum_poiss}, we get
\begin{equation}
    \kappa^{(r_1,...,r_l)}(X_1,...,X_l|Y)= \left\{ \begin{array}{ll} I_{j_k} ~~\textrm{if}~~r_k>0, r_1 = ... = r_{k-1}=r_{k+1}=...r_l =0 \\ 0~~\textrm{in all other cases}
\end{array} \right.
\end{equation}
Therefore, many terms in the total cumulance expansion \eqref{eq:app_total_cumulance} will vanish, and the only non-zero terms are those, in which partition $\pi$ contains only subsets $B$ consisting of one repeating variable. Let $\pi$ be a partition in which $r_1$ variables $X_1$ were divided into $i_1$ subsets, $r_2$ variables $X_2$ were divided into $i_2$ subsets, etc. Then, 
\begin{equation}
    \kappa_Y (\kappa(X_i:i \in B|Y) : B \in \pi) = \kappa^{(i_1, ...,i_l)}(I_{j_1},...,I_{j_l})
\end{equation}
Moreover, there are $S(r_1,i_1)...S(r_l,i_l)$ such partitions $\pi$, where $S(r,i)$ denotest the number of partitions of $r$-element set into $i$ subsets. Therefore,
\begin{equation}
\label{eq:app_cumn_cumI_inv}
    \kappa^{(r_1,...,r_l)} \left(n_{j_1}, ...,n_{j_l}\right) = \sum_{i_1 = 0}^{r_1} \sum_{i_2 = 0}^{r_2} ... \sum_{i_l=0}^{r_l} S(r_1, i_1) ... S(r_l,i_l) \kappa^{(i_1,...,i_l)} \left( I_{j_1},...,I_{j_l} \right).
\end{equation}
Numbers $S(r,i)$ are called Stirling numbers of the 2nd type, and they have the following property \cite{NIST:DLMF}
\begin{equation}
\label{eq:app_stirling_1_stirling_2}
    \sum_{k=0}^{n} s(n,k)\, S(k,m) = \delta_{n,m}= \sum_{k=0}^{n} S(n,k)\, s(k,m)
\end{equation}
where $s(r,i)$ are Stirling numbers of the 1st type, defined as the coefficients in the following expansion
\begin{equation}
    x (x+1)...(x+n-1) = \sum_{k=0}^n s(n,k) x^k
\end{equation}
Equation \eqref{eq:app_stirling_1_stirling_2} allows us to invert relation \eqref{eq:app_cumn_cumI_inv}, which leads to \eqref{eq:cum_n_cum_I} from the main text.
\section{Details of estimator construction}
\label{app:estimator}
Let us remind that vectors  ${\vec \mu}^{(n)}$ and ${ \vec \kappa}^{(I)}$ collect all theoretical photon-count moments and intensity cumulants we want to use;  analogously, vectors $\hat{\vec \mu}^{(n)}$ and $ \hat{ \vec \kappa}^{(I)}$ collect estimators of photon-count moments and estimators of intensity cumulants; elements of $\hat{\vec \mu}^{(n)}$ are given by \eqref{eq:samp_moments}, elements of $ \hat{ \vec \kappa}^{(I)}$ can be expressed in terms of $\hat{\vec \mu}^{(n)}$ using  \eqref{eq:cumn_est} and \eqref{eq:cum_n_cum_I}, and the same equations can be used to express ${ \vec \kappa}^{(I)}$ in terms of ${\vec \mu}^{(n)}$.  We can define a Jacobian matrix of transformation from  ${\vec \mu}^{(n)}$ to $ { \vec \kappa}^{(I)}$ as 
\begin{equation}
      \frac{\partial [{ \vec \kappa}^{(I)} ]_i}{ \partial [{\vec \mu}^{(n)}]_j }=\boldsymbol{J}_{ij} ({\vec \mu}^{(n)}),
\end{equation}
elements of this matrix  can be found using \eqref{eq:cumn_est} and \eqref{eq:cum_n_cum_I}. The matrix $\boldsymbol{J}_{ij} (\hat {\vec \mu}^{(n)})$ is a Jacobian of transformation from $\hat{\vec \mu}^{(n)}$ to  $ \hat{ \vec \kappa}^{(I)}$.
Cumulants estimators calculated using \eqref{eq:cumn_est} and \eqref{eq:cum_n_cum_I} are asymptotically unbiased, which means that
\begin{equation}
\langle \hat {\vec \kappa}^{(I)} \rangle \overset{M \rightarrow \infty}{\longrightarrow} \vec \kappa^{(I)} = \boldsymbol{A} \vec \theta.
\end{equation}
Let us observe that \eqref{eq:samp_moments} can be written as $\hat{\vec \mu}^{(n)} = \sum_{m=1}^M \vec n_m$, where each $\vec n_m$ is an independent random vector containing the products of photon counts 
$n_{j_1,m}\cdots n_{j_r,m}$ used to estimate the moments.  
By the central limit theorem,  
\begin{equation}
    \hat{\vec \mu}^{(n)} \overset{M \rightarrow \infty}{\longrightarrow} 
    \mathcal{N}(\vec \mu^{(n)},\; \boldsymbol{\Sigma}^{(n)}/M),
\end{equation}
where $\mathcal{N}(\vec \mu, \boldsymbol{\Sigma})$ denotes multivariate normal distribution with mean vector $\vec \mu$ and a covariance matrix $\boldsymbol{\Sigma}$. In our case, a covariance matrix can be estimated from collected data using
\begin{equation}
\label{eq:app_cov_n_est}
    \hat{\boldsymbol{\Sigma}}^{(n)} = \frac{1}{M} \left(\sum_{m=1}^M \vec n_m \vec n_m^T\right) - \hat{\vec \mu}^{(n)} \hat{\vec \mu}^{(n) T} 
\end{equation}
Cumulants estimators are functions of moments estimators, so from delta method \cite{Wolter2007},  $\hat{\vec \kappa}$  is also normally distributed, and
\begin{equation}
\label{eq:app_cum_I_model}
    \hat{\vec \kappa}^{(I)} \overset{M \rightarrow \infty}{\longrightarrow} \mathcal{N}\left(\boldsymbol{A} \vec \theta, \boldsymbol{\Sigma}^{(\kappa)}/M\right),~~\boldsymbol{\Sigma}^{(\kappa)} = \boldsymbol{J}(\vec \mu^{(n)}) \boldsymbol{\Sigma}^{(n)} \boldsymbol{J}^T(\vec \mu^{(n)}).
\end{equation}

We proved that cumulants estimators are described by linear Gaussian model \eqref{eq:cum_I_model}. This allows us to construct a simple maximum likelihood estimator \cite{kay1993fundamentals} of object's spatial moments
\begin{equation}
    \hat{\vec \theta} = \left(\boldsymbol{A}^T \left({\boldsymbol{\Sigma}}^{(\kappa)}\right)^{-1} \boldsymbol{A}\right)^{-1} \boldsymbol{A}^T 
    \left({\boldsymbol{\Sigma}}^{(\kappa)}\right)^{-1} \hat{\vec{\kappa}}^{(I)}
\end{equation}
We cannot use this formula directly because we do not know the covariance matrix ${\boldsymbol{\Sigma}}^{(\kappa)}$; however, we can replace ${\boldsymbol{\Sigma}}^{(\kappa)}$ with its estimator constructed as
\begin{equation}
\label{eq:app_Sigmak}
    \hat {\boldsymbol{\Sigma}}_0^{(\kappa)} = \boldsymbol{J}(\hat {\vec \mu}^{(n)}) \hat{\boldsymbol{\Sigma}}^{(n)} \boldsymbol{J}^T(\hat{\vec \mu}^{(n)}),
\end{equation}
where $\hat{\boldsymbol{\Sigma}}^{(n)}$ is constructed using \eqref{eq:app_cov_n_est}, $\hat{\vec \mu}^{(n)}$ entering $\boldsymbol{J}(\hat {\vec \mu}^{(n)})$ are constructed using \eqref{eq:samp_moments}; this leads to initial moments estimate
\begin{equation}
    \hat{\vec \theta}_0 = \left(\boldsymbol{A}^T \left({\hat {\boldsymbol{\Sigma}}_0}^{(\kappa)}\right)^{-1} \boldsymbol{A}\right)^{-1} \boldsymbol{A}^T 
    \left({\hat {\boldsymbol{\Sigma}}_0}^{(\kappa)}\right)^{-1} \hat{\vec{\kappa}}^{(I)}
\end{equation}
We can further increase the estimation accuracy and robustness by using estimated moments to create new, more accurate estimators ${\hat {\boldsymbol{\Sigma}}}^{(\kappa)}$, which are then used to reestimate $\theta$; subsequent estimators can be constructed iteratively:
\begin{equation}
    %{\hat {\boldsymbol{\Sigma}}}^{(\kappa)}_{l+1} = {{\boldsymbol{\Sigma}}}^{(\kappa)}(\hat {\vec \theta}_l), \quad 
    \label{eq:app_iterative_est}
    \hat{\vec \theta}_{l+1} = \left(\boldsymbol{A}^T \left(\hat{\boldsymbol{\Sigma}}^{(\kappa)}_{l+1}(\hat {\vec \theta}_l)\right)^{-1} \boldsymbol{A}\right)^{-1} \boldsymbol{A}^T 
    \left({\hat {\boldsymbol{\Sigma}}_{l+1}}^{(\kappa)}(\hat {\vec \theta}_l)\right)^{-1} \hat{\vec{\kappa}}^{(I)},
\end{equation}
where $\hat{\boldsymbol{\Sigma}}^{(\kappa)}_{l+1}(\hat {\vec \theta}_l)$ means that we calculate covariance matrix estimator assuming $\hat {\vec \theta}_l$ are real moments---firstly, we calculate cumulants as a function of  $\hat {\vec \theta}_l$ using \eqref{eq:cum_I_ser}, then we transform intensity cumulants to photon-count moments using \eqref{eq:app_cumn_cumI_inv} and \eqref{eq:app_cum_to_mom}, and finally, covariance matrix estimator is calculated using \eqref{eq:app_cov_n_est} and \eqref{eq:app_Sigmak}. 

Let us now comment on the effect on the dark-counts noise on the estimator construction and the computation of the C-R bounds. We assume that the dark counts in each detector are independent, Poissonian random variables with mean values  $\mu_\textrm{DC}$. This additional noise affects the equation for theoretical  intensity cumulants \eqref{eq:cum_I_ser} in the following way:  all first-order cumulants (mean detectors signals) are increased by $\mu_\textrm{DC}$, and the remaining, higher-order  intensity cumulants are not affected; this last property comes from cumulants properties \eqref{eq:app_cum_linearity} and \eqref{eq:app_cum_independent}, and the fact that the considered dark-count noise corresponds to additional constant, non-random \textit{intensity} in each detector (additional \textit{photon counts} are random though). Then, one can compute the corresponding photon-count cumulants $\kappa^{(n)}$ using \eqref{eq:app_cumn_cumI_inv}; note that higher-order photon-count cumulants are affected by the dark counts, unlike intensity cumulants. Subsequently, one can compute the corresponding theoretical photon-count moments $\mu^{(n)}$ using \eqref{eq:app_cum_to_mom}, and calculate the FI matrix using \eqref{eq:CR_gaussian}, which allows us to derive C-R bounds in the presence of dark counts. 

The efficient spatial moments estimator in the presence of dark counts must be modified to
\begin{equation}
    \hat{\vec \theta} = \left(\boldsymbol{A}^T \left({\boldsymbol{\Sigma}}^{(\kappa)}\right)^{-1} \boldsymbol{A}\right)^{-1} \boldsymbol{A}^T 
    \left({\boldsymbol{\Sigma}}^{(\kappa)}\right)^{-1} \left( \hat{\vec{\kappa}}^{(I)} -  {\vec{\kappa}}^{(I)}_0(\mu_\textrm{DC})\right),
\end{equation}
where ${\vec{\kappa}}^{(I)}_0(\mu_\textrm{DC})$ contains intensity cumulants values corresponding to zero signal and dark counts of strength $\mu_\textrm{DC}$; elements of this vector corresponding to first order cumulants are equal to $\mu_\textrm{DC}$; other elements are zero. The equation for the initial estimate of a covariance matrix \eqref{eq:app_Sigmak} remains unchanged; in the further iterative estimation procedure given by \eqref{eq:app_iterative_est}, one needs to explicitly include background noise when constructing spatial-moments-based estimation of a covariance matrix $\hat{\boldsymbol{\Sigma}}^{(\kappa)}_{l+1}(\hat {\vec \theta}_l)$. More precisely, one needs to add $\mu_\textrm{DC}$ to all first-order cumulants constructed based on $\hat {\vec \theta}_l$, and perform all remaining reconstruction steps of $\hat{\boldsymbol{\Sigma}}^{(\kappa)}_{l+1}(\hat {\vec \theta}_l)$ computation as described earlier for non-dark-count case.
\section{The role of object size $\Delta$}
\label{app:obj_size}
\begin{figure}
    \centering
    \includegraphics[width = 0.5\columnwidth]{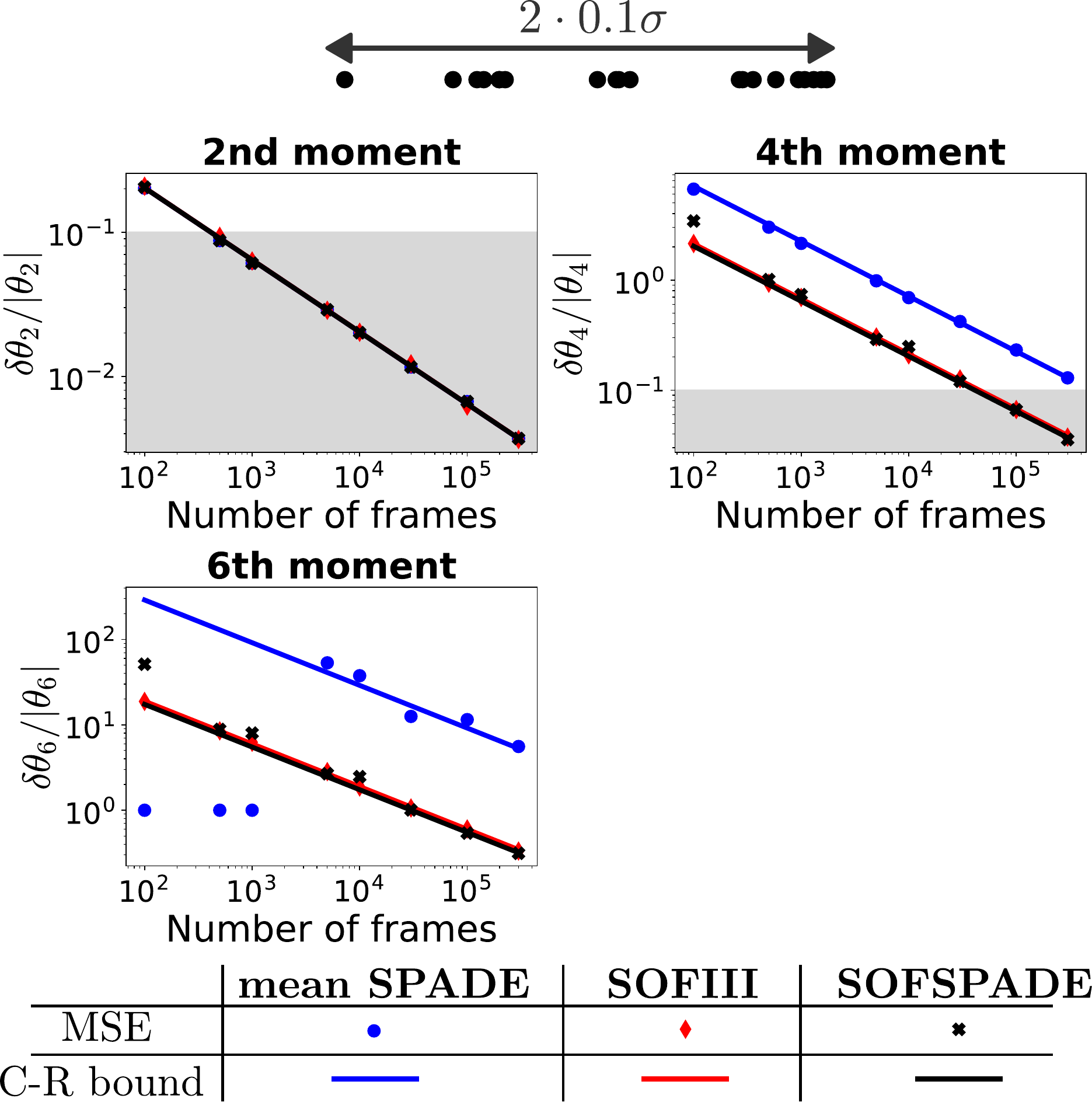}
    \caption{
    The simulated MSEs and corresponding C-R bounds of mean SPADE, SOFSPADE, and SOFIII are plotted as functions of the number of frames, for the estimation of even moments of the depicted subdiffraction object of size $\Delta = 0.1$. This object is smaller than the one considered in the main text, so the estimation MSEs are larger. For the 6th moment and low frame numbers, one can notice that the MSE falls below the C-R bound; this is, however, a result of the estimator's bias---the signal is so low that the 6th moment is estimated to be $0$, leading to a relative error of $1$.}
    \label{fig:app_SPADE_plots_0_1}
\end{figure}

The simulations in the main text were performed for a fixed normalized object size $\Delta = 0.3$. Let us now perform analogous simulations of even-moment reconstruction for smaller ($\Delta = 0.1$) and larger ($\Delta = 0.8$) objects, created by rescaling the object depicted in Fig.~3a.
For $\Delta = 0.1$, we observe that the 8th moment is very close to 0: with the number of frames used in our simulations, it is practically impossible to estimate the 8th moment for such a small object, and it is reasonable to neglect this moment and estimate only the lower even moments, $\vec \theta = [\theta_0,\theta_2,\theta_4,\theta_6]$. Then, the set of cumulants used to reconstruct these moments with mean SPADE is $\hat {\vec \kappa} = [ \hat \kappa(I_0^\textrm{S}),\hat \kappa(I_1^\textrm{S}),\hat \kappa(I_2^\textrm{S}),\hat \kappa(I_3^\textrm{S})]^T$ (note that 4 detectors are sufficient), the analogous set for SOFIII is $\hat {\vec \kappa} = [\hat{\kappa}(I_+^\textrm{III}), \hat{\kappa}(I_-^\textrm{III}), \hat{\kappa}^{(2)}(I_-^\textrm{III}), \hat{\kappa}^{(3)}(I_-^\textrm{III})] $, and for SOFSPADE
$$
    \hat {\vec \kappa} = [ \hat \kappa(I_0^\textrm{S}),\hat \kappa(I_1^\textrm{S}),\hat \kappa(I_2^\textrm{S}),\hat \kappa(I_3^\textrm{S}),\hat \kappa^{(2)}(I_1^\textrm{S}),\hat \kappa(I_1^\textrm{S}, I_2^\textrm{S}), \hat \kappa^{(3)}(I_1^\textrm{S}) ]^T.
$$
The MSEs of the moment estimates and corresponding C-R bounds for the considered object with $\Delta = 0.1$ are shown in Fig.~\ref{fig:app_SPADE_plots_0_1}. The blinking parameters are the same as those used in the main text.

\begin{figure}
    \centering
    \includegraphics[width = 0.5\columnwidth]{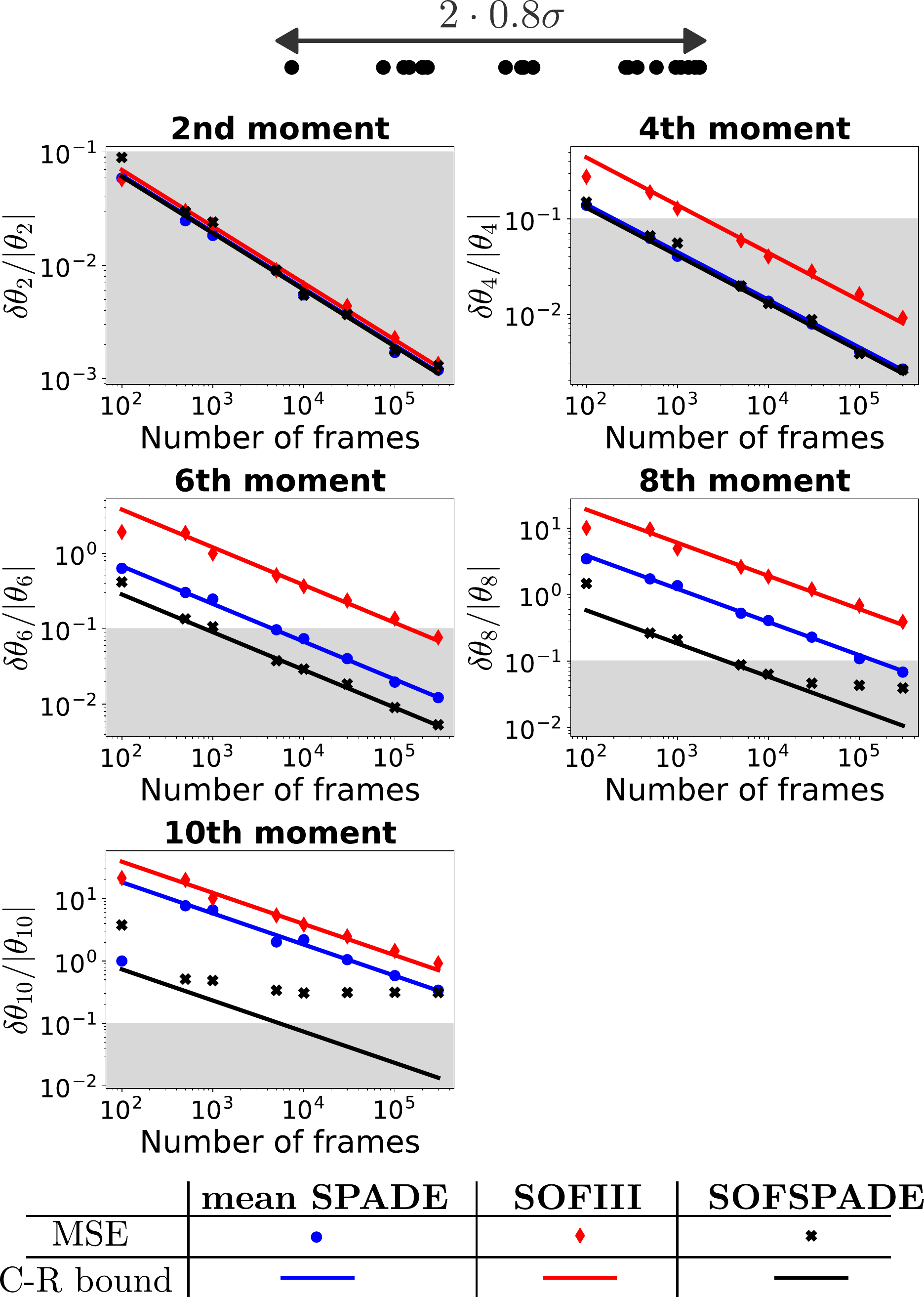}
    \caption{
    The simulated MSEs and corresponding C-R bounds of mean SPADE, SOFSPADE, and SOFIII are plotted as functions of the number of frames, for the estimation of even moments of the depicted object of size $\Delta = 0.8$. Moments up to order 10 are estimated, but higher-order moments cannot be entirely neglected; this is manifested by systematic errors in the 8th- and 10th-moment estimates obtained with SOFSPADE, which cannot be reduced by increasing the number of frames.}
    \label{fig:app_SPADE_plots_0_8}
\end{figure}

On the contrary, for $\Delta = 0.8$, higher-order moments play an important role, as they do not decay so quickly with the moment order. Estimating too few of these moments introduces systematic errors when non-negligible terms are truncated from the series expansion in \eqref{eq:cum_I_ser}. Therefore, we perform the estimation of even moments up to 10th order, $\vec \theta = [\theta_0, \theta_2, \theta_4, \theta_6, \theta_8, \theta_{10}]^T$, and the cumulant sets used for mean SPADE, SOFIII, and SOFSPADE are extended accordingly. The estimation errors and fundamental bounds for this case are depicted in Fig.~\ref{fig:app_SPADE_plots_0_8}.

One can notice that the MSEs of the SOFSPADE estimators for $\theta_8$ and $\theta_{10}$ do not decrease with the number of frames below a certain value. This is the sign of a systematic estimation error, which cannot be decreased by increasing the number of frames. This systematic error is caused by a non-negligible value of moments of order 12 and higher. This problem can be fixed by adding these higher moments to the estimation procedure, which would, however, increase the estimation complexity.

One should also notice that the information carried by cumulants of lower-order H-G modes is more relevant for small $\Delta$ values. For $\Delta = 0.1$, SOFIII and SOFSPADE perform almost equally well, both significantly better than mean SPADE. For $\Delta = 0.8$, SOFSPADE still provides the best estimation precision, but SOFIII performs worse than mean SPADE, which indicates that the mean signal of higher-order H-G modes is more informative than the cumulants of lower-order H-G modes.

At this point, it is natural to ask about the scaling of the estimation precision of different moments with object size $\Delta$---this scaling is examined in Sec.~\ref{app:theory}.\ref{sec:app_mom_theo}.
\section{Theoretical aspects}
\label{app:theory}
This section reviews the fundamental limitations of established superresolution techniques such as SOFI and SPADE, and discusses theoretical bounds for the newly introduced SOFSPADE and SOFIII methods. 
\subsection{Two points separation}
 Let us begin by studying the simple scenario, in which the only unknown parameter is the separation $s$ between two identical point emitters, see Fig.~\ref{fig:theo_tab}, 3rd column. For DI with $n$ available photons, the MSE of optimal estimator of $s$ satisfies, for small $s$, $\delta \hat s \sqrt{n}= 2 \sqrt2 \sigma^2/s + \mathcal{O}(1)$, so it diverges for $s \rightarrow \infty$ \cite{Tsang_2016}. This can be remedied  by SPADE technique, for which $\delta \hat s \sqrt{n}= 0.25 \sigma$, so the MSE is constant independently of $n$. A similar improvement is achievable with III, for which $\delta \hat s \sqrt{n}= 0.25 \sigma + \mathcal{O}(s^2)$; those two techniques are equivalent for $s \rightarrow 0$, but SPADE outperforms III for larger $s$.

 The improvement coming from SOFI is different---the MSE still diverges for $s \rightarrow 0$, but fluctuations allow us to decrease this MSE by some constant depending on blinking dynamics\cite{kurdzialek2021super}, $\delta \hat s \sqrt{n}= \beta \sigma^2/s + \mathcal{O}(1)$, $\beta < 2 \sqrt{2}$, see Fig.~\ref{fig:theo_tab}.

 Interestingly, fluctuations do not allow for further precision improvement of estimation of $s$ for SPADE and III measurements; for this task, mean SPADE is equivalent to SOFSPADE, and III is equivalent to SOFIII. Moreover, the fundamental limit for separation estimation precision achievable using \emph{arbitrary} measurement allowed by quantum mechanics is the same whether fluctuations are present or not, and SPADE measurement saturates this limit in both fluctuating and non-fluctuating cases. 

Let us prove the above statements. 

For two sources in the object plane placed at positions $x_+ = s/2$, $x_- = -s/2$, the single-photon quantum state observed in the image plane is

\begin{equation}
\label{eq:rho_single_phot}
    \rho_s = \frac{1}{2} \left( \ket{u_{+}} \!\!\bra{u_{+}} + \ket{u_{-}}\!\! \bra{u_{-}}\right),
\end{equation}
where 
\begin{equation}
\label{eq:udef}
    \ket{u_\pm} = \int \textrm{d} x' u(x' - x_\pm) \ket{x'}
\end{equation}
are  pure states of photons emitted from right/left source respectively, with $u(x')$ satisfying $\left|u(x')\right|^2 = U(x')$, $\ket{x'}$ are eigenstates of a position operator. 

For non-fluctuating sources, subsequent photons are not correlated, so the state of $n$ photons is
\begin{equation}
    \rho_s^{(n)} = \rho_s^{\otimes n}
\end{equation}
The MSE of  estimation of $s$ is fundamentally lower-bounded by a quantum Cram\'er-Rao bound,
\begin{equation}
    \delta \hat s \ge \frac{1}{F_Q(\rho_s^{(n)})},
\end{equation}
where $F_Q$ is the Quantum Fisher Information (QFI) \cite{holevo2011probabilistic}, which is classical Fisher Information maximized over all measurement that can be performed on $\rho_s^{(n)}$. The QFI for an arbitrary $s$-dependent quantum state $\rho_s$ can be calculated as
\begin{equation}
    F_Q(\rho_s) = \textrm{Tr}(\rho_s L_s^2), \quad 
    \partial_s \rho_s = \frac{1}{2} \left( \rho_s L_s + L_s \rho_s \right),
\end{equation}
where the second part of the above equation is the definition of $L_s$, which is called symmetric logarithmic derivative (SLD) matrix.

For non-fluctuating sources and Gaussian PSF, it was shown that \cite{Tsang_2016}
\begin{equation}
    F_Q(\rho_s^{(n)}) = n F_Q(\rho_s) = \frac{n}{4 \sigma^2},
\end{equation}
where QFI additivity with respect to tensor product was used.
Moreover, it was shown that the quantum Cram\'er-Rao bound can be saturated using a measurement in H-G modes $\ket{\phi_j}$, whose wave functions in a position basis are given by \eqref{eq:HG_modes}. 

Let $r_i \in \{+,-\}$ denote the position of the emitter from which the $i$th photon was emitted ($+$ for $x_+$, $-$ for $x_-$). We treat photons as distinguishable, which is justified in the weak-source regime where temporal modes of different photons are orthogonal---this assumption breaks down, for example, for strong thermal sources \cite{Nair2016PRL}. In the absence of correlations, each photon comes from the $x_+$ or $x_-$ source with equal probability $1/2$, so the total probability that the $i$th photon comes from source $r_i$ for $i \in \{1,2,\ldots,n\}$ is
$P(r_1,\ldots,r_n) = \frac{1}{2^n}$
for any $r_1, r_2, \ldots, r_n$.

Let us now consider a general correlated model for which the multi-variable probability distribution $P(r_1,\ldots,r_n)$ can be arbitrary. The density matrix of $n$ photons is then
\begin{equation}
    \rho_s^{(n)} = \sum_{r_1,\ldots,r_n \in \{+,-\}} P(r_1,r_2,\ldots,r_n) \bigotimes_{i=1}^n \ket{u_{r_i}}\!\!\bra{u_{r_i}}.
\end{equation}
It is challenging to compute the QFI of $\rho_s^{(n)}$, as it cannot generally be written as a tensor product of single-photon density matrices. However, an upper bound can be found using QFI convexity \cite{holevo2011probabilistic}, from which we obtain
\begin{equation}
\label{eq:QFI_k_bound_init}
    F_Q\left(  \rho_s^{(n)} \right) \le \sum_{r_1,\ldots,r_n \in \{+,-\}} P(r_1,r_2,\ldots,r_n) F_Q \left( \bigotimes_{i=1}^n \ket{u_{r_i}}\!\!\bra{u_{r_i}} \right).
\end{equation}
The QFI of a pure state $\ket{u_{r_i}}\!\!\bra{u_{r_i}}$ can be calculated directly, yielding
\begin{equation}
\label{eq:QFI_u}
    F_Q \left( \ket{u_{r_i}}\!\! \bra{u_{r_i}} \right) = \frac{1}{4 \sigma^2},
\end{equation}
which is the QFI of a single source localization. Using QFI additivity, inserting \eqref{eq:QFI_u} into \eqref{eq:QFI_k_bound_init}, and using the fact that probabilities $P(r_1,r_2,\ldots,r_n)$ sum to $1$, we obtain the QFI upper bound
\begin{equation}
\label{eq:QFI_upp_bound}
    F_Q\left(  \rho_s^{(n)} \right) \le \frac{n}{4 \sigma^2},
\end{equation}
which is sufficient to show that correlations do not increase the QFI.

Interestingly, they also do not \textit{decrease} it. To prove this, note that for a measurement in H-G modes \eqref{eq:HG_modes} we have
\begin{equation}
    |\braket{\phi_j|u_+}|^2 = |\braket{\phi_j|u_-}|^2
\end{equation}
for any $j$ because $\phi_j(x')$ are either even or odd functions of $x'$. Consequently, if we measure each photon independently in the H-G basis, the statistics of the measurement results are not affected by the choice of $P(r_1,r_2,\ldots,r_n)$, since it does not matter from which source a photon was emitted. This means that the classical FI associated with this measurement applied to $n$ photons is $F = \frac{n}{4 \sigma^2}$, which is the result obtained for the uncorrelated case, where the H-G measurement is known to be QCRB-saturating. However, the classical FI of any measurement always provides a \textit{lower bound} for the QFI, so we can conclude that
$$F_Q\left(  \rho_s^{(n)} \right) \ge \frac{n}{4 \sigma^2},$$
which together with \eqref{eq:QFI_upp_bound} leads to the conclusion that
\begin{equation}
    F_Q\left(  \rho_s^{(n)} \right) = \frac{n}{4 \sigma^2}
\end{equation}
for any distribution $P(r_1,\ldots,r_n)$, and that H-G mode measurement is optimal regardless of the correlation type.

 \subsection{Moments estimation for complex objects}
\label{sec:app_mom_theo}
Let us now discuss a more  complicated task of estimation of spatial moments $\theta_\mu$ of arbitrary incoherent objects.

As mentioned earlier, the MSE of $\mu$th spatial moment for non-fluctuating sources  satisfies $\delta \hat \theta_\mu / \hat \theta_\mu = \mathcal{O}(\Delta^{-\mu})$ for DI and $\delta \hat \theta_\mu / \hat \theta_\mu = \mathcal{O}(\Delta^{-\lceil \frac{\mu}{2}\rceil})$ for SPADE---moreover, SPADE provides optimal precision scaling among all possible measurements.

\begin{figure}
    \centering
    \includegraphics[width = 0.8\columnwidth]{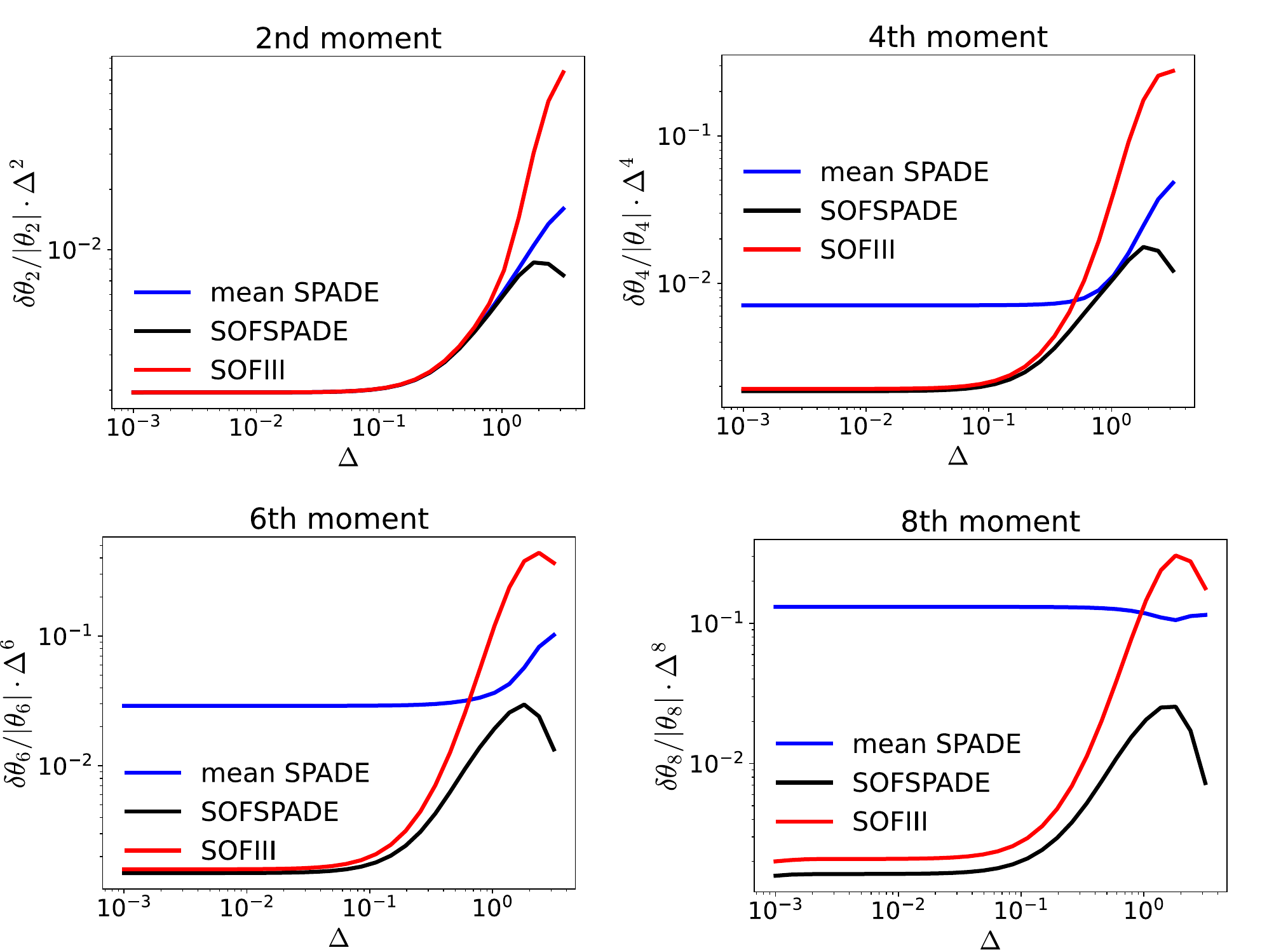}
    \caption{Cram\'er-Rao bound for relative MSE of moments estimation as a function of an object size $\Delta$. In numerical simulations, the same object as in the main text (Fig. 3) was used, rescaled by an appropriate factor. One can see that $\delta \theta_\mu / |\theta_\mu| \cdot \Delta^{\lceil \frac{\mu}{2}\rceil}$ becomes constant for sufficiently small $\Delta$, which is the numerical evidence for the MSE scaling.  The same behavior was observed for odd and even  moments estimation using iSPADE / binary SOFiSPADE. One can also observe that the scaling constant becomes smaller when fluctuations are used for estimation. The data shown are computed for $M=10^4$ frames, but the change of number of frames would only rescale the $y$ axis values.  }
    \label{fig:scaling_even}
\end{figure}

\begin{figure}
    \centering
    \includegraphics[width = 0.7\columnwidth]{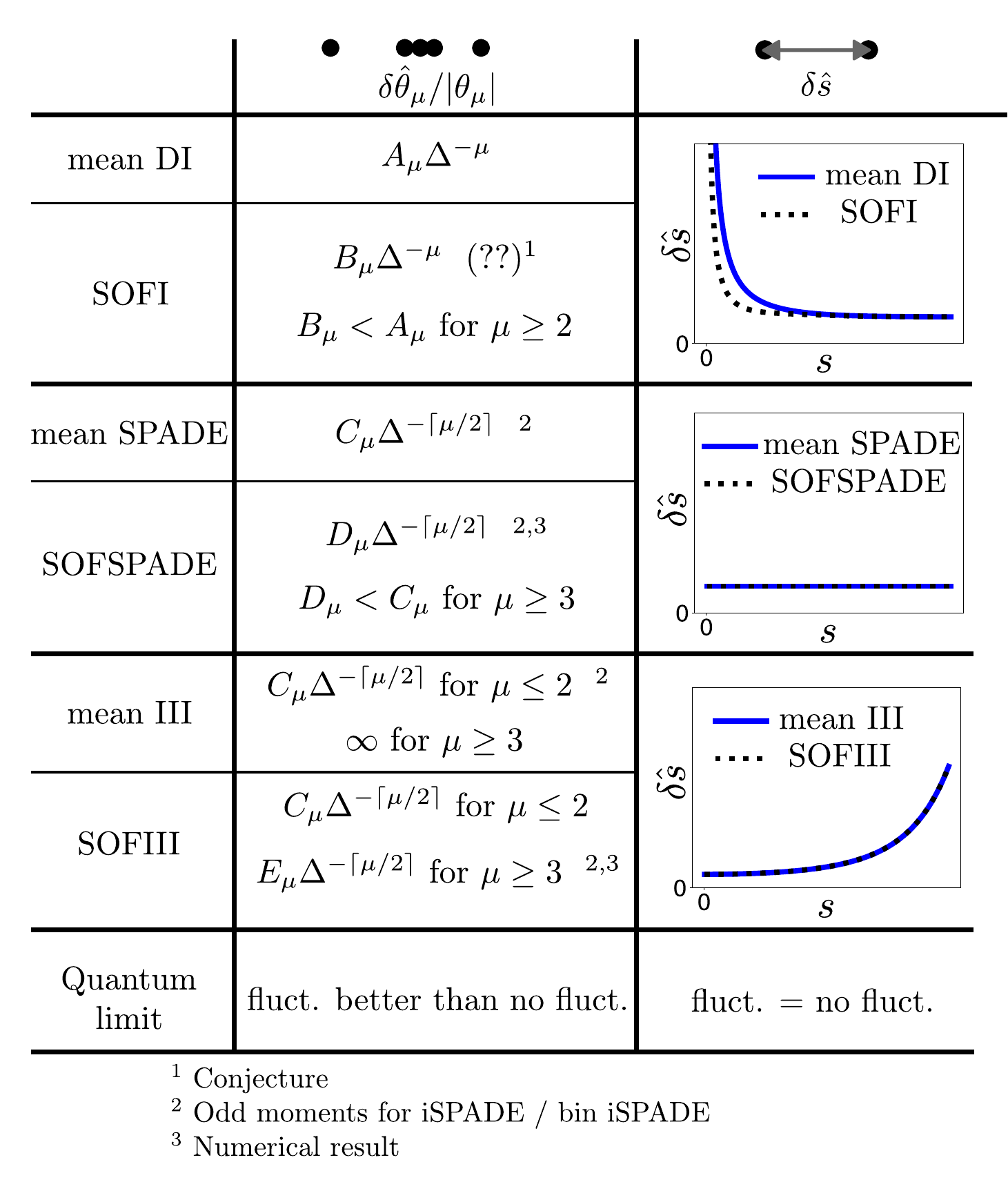}
    \caption{Summary of theoretical limits for different imaging techniques. For moments estimation task, SPADE measurement offers better precision scaling with object size $\Delta$ than direct imaging. The scaling coefficient (but not exponent) can be further improved by utilizing fluctuations. For the estimation of separation between two equally bright point emitters, fluctuations help to improve the precision of DI measurement, but do not give any advantage, when optimal (SPADE) measurement is chosen.}
    \label{fig:theo_tab}
\end{figure}

Our results show that higher moments estimation precision can be improved using fluctuations. As we observed numerically (see Fig.~\ref{fig:scaling_even}), the scaling $\delta \hat \theta_\mu / \hat \theta_\mu = \mathcal{O}(\Delta^{-\lceil \frac{\mu}{2}\rceil})$ remains true for SOFSPADE---however, fluctuations allow us to improve a \textit{constant} in this scaling. Moreover, the same scaling is achievable for SOFIII. The crucial difference between mean III and SOFIII is that the latter technique recovers full information about the object.

The fundamental precision limits for different imaging techniques are summarized in Fig.~\ref{fig:theo_tab}. The optimal performance of SOFI for moments estimation of complex objects is not known---however, we conjecture that the precision scaling exponent is the same for DI and SOFI---the SOFI technique only makes the scaling constant better. This conjecture is based on the analogous relation between SPADE and SOFSPADE, observed numerically. Moreover, this conjecture is also supported by the behavior of SOFI precision for two point sources separation estimation. Unlike for SPADE, SOFI does not make $\delta \hat s$ finite for $s \rightarrow 0$, which suggests only a constant improvement for moments estimation tasks.
%\label{sec:theoretical}

%\bibliographystyle{unsrt}

\end{document}